\documentclass[10pt,journal]{IEEEtran}
\usepackage[latin9]{inputenc}
\usepackage{array}
\usepackage{bm}
\usepackage{multirow}
\usepackage{amsmath}
\usepackage{amssymb}
\usepackage{graphicx}
\makeatletter
\providecommand{\tabularnewline}{\\}

\usepackage{epsfig}\usepackage{subfigure}\usepackage{amsfonts}\usepackage{cite}\usepackage{color}\usepackage{bm}\usepackage{booktabs}
\usepackage{multirow}\usepackage{comment}\usepackage{epstopdf}

\newcounter{TempEqCnt}
\makeatother
\begin{document}
\title{An Augmented Nonlinear LMS for Digital Self-Interference
Cancellation in Full-Duplex Direct-Conversion Transceivers}
\author{Zhe Li, \emph{Student Member}, \emph{IEEE}, Yili Xia, \emph{Member},
\emph{IEEE}, Wenjiang Pei, Kai Wang, and Danilo P. Mandic, \emph{Fellow},
\emph{IEEE} 
\thanks{This work was partially supported by the National Natural Science
Foundation of China under Grant 61401094 and Grant 61771124, the Natural Science Foundation
of Jiangsu Province under Grant BK20140645, the Fundamental Research
Funds for the Central Universities under Grant 2242016K41050, and the
China Scholarship Council. (\emph{Corresponding authors: Yili Xia; Wenjiang Pei.})} \thanks{Z. Li is with the School of Information Science and Engineering, Southeast
University, 2 Sipailou, Nanjing 210096, P. R. China, and also with
the Department of Electrical and Electronic Engineering, Imperial
College London, London SW7 2AZ, U.K. (e-mail:lizhe\_nanjing@seu.edu.cn)} \thanks{Y. Xia, W. Pei, and K. Wang are with the School of Information Science
and Engineering, Southeast University, 2 Sipailou, Nanjing 210096,
P. R. China. (e-mail: yili\_xia@seu.edu.cn; wjpei@seu.edu.cn; kaiwang@seu.edu.cn).} \thanks{D. P. Mandic is with the Department of Electrical and Electronic Engineering,
Imperial College London, London SW7 2AZ, U.K. (e-mail: d.mandic@imperial.ac.uk).} }
\maketitle
\begin{abstract}
In future full-duplex communications, the cancellation of self-interference (SI) arising from hardware non-idealities will play an important role in the design of mobile-scale devices. To this end, we introduce an
optimal digital SI cancellation solution for shared-antenna-based direct-conversion transceivers. To establish that the underlying widely linear signal model is not adequate for strong transmit signals, the impact of various circuit imperfections, including power amplifier (PA) distortion, frequency-dependent I/Q imbalance, quantization noise and thermal noise, on the performance of the conventional augmented least mean square (LMS) based SI canceller, is analyzed. In order to achieve a sufficient signal-to-interference-plus-noise ratio (SINR) when the nonlinear SI components are not negligible, we propose an augmented nonlinear LMS based SI canceller for a joint cancellation of both the linear and nonlinear SI components by virtue of a widely nonlinear model fit. A rigorous mean and mean square performance evaluation is conducted to justify the performance advantages of the proposed scheme over the conventional  augmented LMS solution. Simulations on orthogonal frequency division multiplexing (OFDM)-based wireless local area network (WLAN) standard compliant waveforms support the analysis.
\end{abstract}
\IEEEpeerreviewmaketitle{ }
\vspace{-0.3cm}
\section{Introduction}
\IEEEPARstart{T}{he} full-duplex (FD) technology aims at doubling the radio link data rate through simultaneous and bidirectional communication at the same center frequency, and is widely considered as a driving-force behind more spectrally efficient wireless networks and a potential candidate to fulfill the ambition of 5G to reach a 1000-fold gain in capacity \cite{Hong2014,Goyal2015}. One of the major challenges in FD communications is the so-called self-interference (SI) problem, that is, a strong transmit
signal coupled into the receiver (Rx) path. Since the transmitter (Tx) and Rx chains are closely linked together in each transceiver node of FD communication systems, the SI power leaked into and reflected from the Tx chain could be even 50 dB to 110 dB higher than the Rx sensitivity level in either wireless local area network (WLAN) or cellular scenarios \cite{Kim2015,Sabharwal2014,Kolodziej2016}. The design of FD transceivers has long been considered impossible for practical realizations and implementations, and it is only recently that their feasibility was experimentally demonstrated using the wireless open-access research platform (WARP) with WiFi waveforms \cite{Choi2010,Jain2011,Duarte2012,Bharadia2013,Bharadia2014,Everett2014}. Based on this promising result, it was recently suggested that a preferable FD network should consist of backhaul nodes operating in the FD mode and access nodes remaining in the legacy half-duplex (HD) mode \cite{Everett2011}. However, recent studies have showed that operating access nodes in the FD mode significantly leverages the gain in degrees of freedom in either ergodic or fast-fading channel \cite{Sahai2013}, and an imperative is to design a hardware structure suitable for mass-production. Owing to the physical constraints, such as small-size, low-cost and low-energy-consumption, direct-conversion transceivers  are widely applied in HD wireless systems, and are also suitable for far-end device implementation in the context of FD communication systems.

In order to provide efficient SI cancellation, there exist numerous types of hardware solutions. According to the antenna placement strategies, these can be classified into separate-antennas-based \cite{Snow2011,Riihonen2011} and shared-antenna-based schemes \cite{Bharadia2013,Bharadia2014}. When each transceiver node is equipped with more than two separate antennas, SI attenuation can be achieved by improving electromagnetic insulation between the antennas. Owing to the inherent closed-loop within FD systems, the knowledge of the SI channel matrix can be obtained by either placing extra transmit antennas or allocating specific spatial resources \cite{Snow2011,Riihonen2011}. On the other hand, the shared-antenna-based design aims to separate the transmit and receive signals by sharing a common antenna \cite{Bharadia2013,Bharadia2014}, the key component of which is a three port routing device, known as a circulator, used to isolate the incoming and outgoing signals. Requiring only off-the-shelf radio-frequency (RF) components, the shared-antenna structure stands out as a cost-effective and energy-saving choice for the design of mobile-scale FD transceivers. Demonstrations on the WARP have shown that in this way even 110 dB and 103 dB SI cancellation can be achieved in SISO \cite{Bharadia2013} and MIMO systems respectively\cite{Bharadia2014}.

\indent In a shared-antenna structure, it is necessary to consider further non-trivial analog and digital SI cancellation, due to the leakage of the circulator, single-path reflection from the antenna, and multi-path interference from the surrounding environment \cite{Korpi2016a}. The purpose of analog SI cancellation is to prevent the saturation of the SI power level within the Rx low-noise amplifier (LNA), and meanwhile, to ensure that the difference between the power of residual SI and the received signal of interest does not exceed the dynamic range of an analog-to-digital converter (ADC) \cite{Choi2013}. Subsequently, further digital cancellation is performed to deal with the residual SI components, as well as other RF circuit non-idealities, mainly including nonlinear distortion, I/Q imbalance and phase noise. The nonlinearity is largely caused by the power amplifier (PA), while I/Q imbalance and phase noise are mainly induced by the imperfect local oscillator (LO). The impact of PA nonlinear distortion on FD transceivers has been investigated in \cite{Anttila2013,Korpi2014a}, while the effect of phase noise was analyzed in \cite{Syrjala2014,Quan2017}. Since the I/Q imbalance is essentially reflected in the mismatch between in-phase and quadrature components of the complex-valued I/Q signal, it is also reflected in an image interference associated with the original signal \cite{Tarighat_2005,Anttila2008a,Anttila_2008,Li2017}.

The impact of the image interference caused by Tx I/Q imbalance on the SI cancellation has been studied in \cite{Li2014}, indicating that it heavily limits the receiver path signal-to-interference-plus-noise ratio (SINR). However, due to size constraints of FD transceivers, the Rx and Tx are required to share a common imperfect LO, therefore, a more accurate analysis of image interference on SI cancellation should be performed by a joint consideration of both Tx and Rx I/Q imbalance.  Motivated by this finding, a widely linear processing framework was developed in \cite{Korpi2014} to jointly suppress both the original transmit SI signal and its complex conjugate, i.e., the image interference, which arises due to the frequency-dependent I/Q imbalance in both the Tx and Rx. The cancellation parameters were subsequently estimated in the widely linear least squares sense. Although the model analysis in \cite{Korpi2014} has illustrated that under certain circumstances, e.g., for large transmit powers, the PA within the FD transceiver is likely to be operating close to or within its saturation region, consequently introducing third order nonlinear distortion, for mathematical simplicity, the higher order SI components have not been considered by the block-based SI canceller. A more general arbitrary nonlinear order PA and a simplified frequency-independent I/Q modulator were considered in \cite{Korpi2017}, and the corresponding nonlinear SI cancellation was addressed by a model-fit widely nonlinear least squares approach.

By exploiting the advantages of adaptive estimation algorithms over block-based least squares ones, such as their lower computational complexities and faster adaptation for potential time-varying channels, the augmented (widely linear) least mean square (LMS) adaptive filtering algorithm \cite{Javidi2008,Mandic_Book2,Xia2012} has been employed in a DSP-assisted analog SI cancellation process, and its theoretical performance in the presence of Tx and Rx IQ imbalance has been evaluated in \cite{Sakai2016a}. For simplicity, the I/Q imbalance within transmitters and receivers were also considered to be frequency-independent in \cite{Sakai2016a}. This is, however, not the case in wideband scenarios, since their frequency selectivity has been extensively reported and justified in \cite{Anttila2008a,Anttila_2008}. Furthermore, although it has been illustrated by simulations that due to the undermodeling problem,  the augmented LMS yields suboptimal SI cancellation results in the presence of PA nonlinearity, while a theoretical understanding of this suboptimality and ways of its mitigation are still lacking.

Therefore, in this paper, we first conduct a comprehensive mean and mean square performance analysis on the augmented LMS based SI canceller for wideband FD transceivers in the presence of both the PA nonlinear distortion and frequency-dependent image interference, to theoretically quantify its bias and variance increase, in the steady-state stage. The proposed analysis consequently facilitates a physical verification on how the augmented LMS SI canceller fails to achieve the required signal-to-noise ratio (SNR) when the transmit power is high enough. Next, in order to achieve a sufficient amount of SINR when the nonlinear SI components are not negligible, an augmented nonlinear LMS, whose underlying estimation framework generally takes into account both the nonlinear SI component and its associated image interference, is proposed for unbiased nonlinear SI cancellation, and a theoretical performance evaluation is conducted to demonstrate its performance advantages over the conventional augmented LMS. It is important to note that such further theoretical performance analysis is not an incremental step from the conventional augmented LMS, mainly due to the non-Gaussian nature of the higher order SI components in wideband scenarios. From the statistical perspective, this analysis also provides physical insights to the evaluation of the theoretical performance bounds on those block-based SI cancellers proposed in \cite{Korpi2014,Korpi2017}. Moreover, to facilitate the use of the proposed augmented nonlinear LMS based SI canceller in practical applications, a data pre-whitening scheme is employed to speed up its convergence. Simulations on orthogonal frequency division multiplexing (OFDM)-based WLAN standard compliant waveforms support the analysis.

\emph{Notations}: Lowercase letters are used to denote scalars, $a$,
boldface letters for column vectors, \textbf{a}, and boldface uppercase
letters for matrices, \textbf{A}. The symbols $\textbf{0}_{N}$ and $\textbf{1}_{N}$ denote respectively an $N\times 1$ zero and unity vector. An $N\times N$ identity matrix is denoted by $\textbf{I}_{N}$. The superscripts $(\cdot)^{*}$, $(\cdot)^{T}$, $(\cdot)^{H}$ and $(\cdot)^{-1}$ denote respectively the complex conjugation, transpose, Hermitian transpose and matrix
inversion operation. The operator $\text{Tr}[\cdot]$ represents the trace
of a matrix, while the operators $\otimes$, $\left\Vert \cdot\right\Vert$ respectively denote the Kronecker product and Euclidean norm. The statistical expectation operator is denoted by $E[\cdot]$, matrix
determinant by $\det[\cdot]$, while the operators $\Re[\cdot]$
and $\Im[\cdot]$ extract respectively the real and imaginary
part of a complex variable and $j=\sqrt{-1}$.
Matrix vectorization is designated by ${\rm {vec}}\{\cdot\}$,
which returns a column vector transformed by stacking the successive
columns of matrix, and its inverse
operation, i.e., restoring the matrix from the its vectorized form,
is denoted by ${{\rm {vec}}^{-1}}\{\cdot\}$. The extraction of matrix diagonal elements into a vector is denoted by $\rm{diag}\{\cdot\}$. The operator ${\lambda_{\max}}[\cdot]$ returns the largest positive eigenvalue of a matrix.
%
\begin{figure*}[t!]
\centering \includegraphics[width=0.8\textwidth]{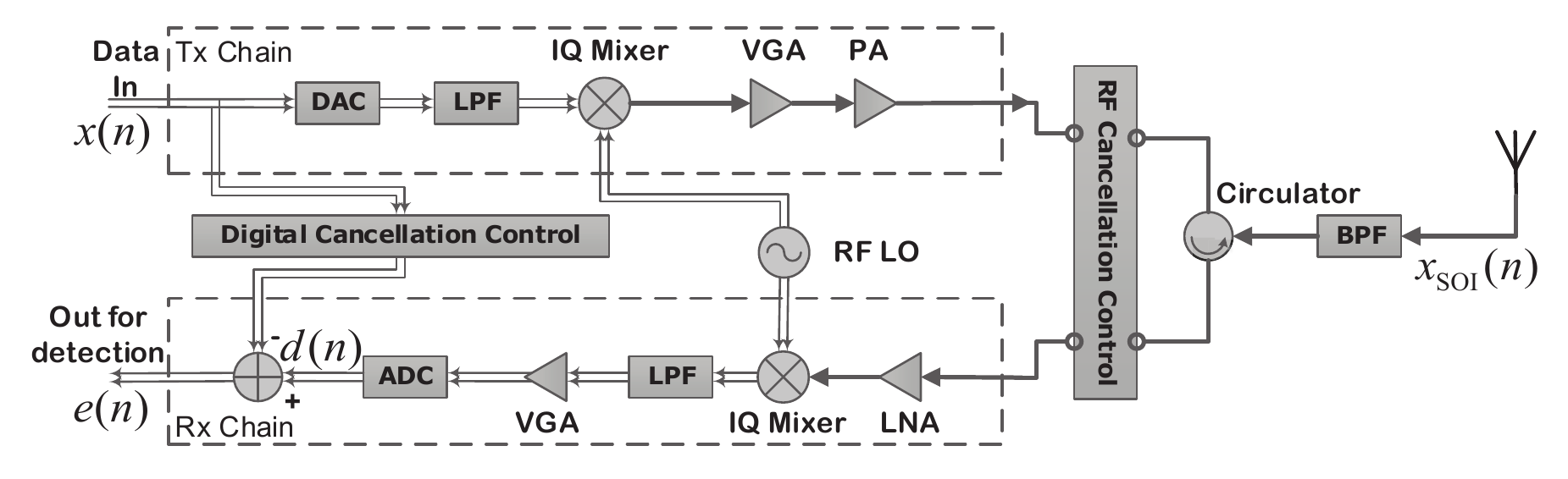}
\caption{The architecture of a shared-antenna FD transceiver.}
\label{fig1}
\end{figure*}
\section{Full-Duplex Transceiver and Its Widely Linear Baseband Equivalent Model}
\label{model}
%
%

The structure of a typical shared-antenna FD direct-conversion transceiver  is given in Fig. \ref{fig1}, and this structure is widely adopted in modern wireless transceivers, due to its simplicity \cite{Mak_2007}. In such an architecture, the leakage of the circulator, the single-path reflection from the antenna, and the multi-path interference from surrounding environment introduce plenty of residual self-interference (SI), which is firstly mitigated by the RF cancellation module and then suppressed by the digital baseband SI canceller \cite{Korpi2014,Korpi2014a}. By considering the fact that low-cost RF components are preferable for the built-up of mobile transceivers, a precise baseband-equivalent system model that incorporates their prominent hardware non-idealities, such as Tx and Rx I/Q imbalance, PA distortion, ADC quantization noise and thermal noise, is a prerequisite for the digital SI cancellation control. This in fact yields a widely linear relation, which at a time instant $n$ during the digital cancellation process, and between the observed signal $d(n)$ at the input of the canceller and its corresponding SI waveform $x(n)$, has the form \cite{Korpi2014}
\begin{equation}
d(n)={{\textbf{x}}^{T}}(n){{\textbf{h}}^{\textmd{o}}}+{{\textbf{x}}^{H}}(n){{\textbf{g}}^{\textmd{o}}}+u(n)\label{dn_vec}
\end{equation}
where ${\textbf{x}}(n) ={[x(n),x(n-1),\ldots,x(n-M+1)]^{T}}$ is an SI vector of length $M$ and is perfectly known by the receiver. The end-to-end channel impulse responses ${{\textbf{h}}^{\textmd{o}}}={[{h_{1}^{\textmd{o}}}, {h_{2}^{\textmd{o}}}, \ldots, {h_{M}^{\textmd{o}}}]^{T}}$ and ${{\textbf{g}}^{\textmd{o}}}={[{g_{1}^{\textmd{o}}}, {g_{2}^{\textmd{o}}}, \ldots, {g_{M}^{\textmd{o}}}]^{T}}$ are determined by frequency-dependent I/Q imbalance in both the transmitter and receiver, PA memory, and residual of analog cancellation. The composite noise term $u(n)$ represents the sum of interference components, including PA nonlinearity, thermal noise and quantization noise from an ADC,  and is given by \cite{Korpi2014}
\begin{equation}
u(n)={\textbf{x}}_{{\rm {IMD}}}^{T}(n){{\textbf{h}}_{{\rm {IMD}}}^{\textmd{o}}}+{\textbf{x}}_{{\rm {IMD}}}^{H}(n){{\textbf{g}}_{{\rm {IMD}}}^{\textmd{o}}}+v(n)+q(n)\label{un_vec}
\end{equation}
where ${{\textbf{x}}_{{\rm {IMD}}}}(n) = [{x_{{\rm {IMD}}}}(n), {x_{{\rm {IMD}}}}(n-1), \dots, {x_{{\rm {IMD}}}}(n-N+1)]^T$ and ${x}_{{\rm {IMD}}}(n)$ represents third-order intermodulation (IMD) SI introduced by PA distortion, given by \cite{Korpi2014}
\begin{equation}
{x_{{\rm {IMD}}}}(n)=k_{{\rm {TIQ}}}^{3/2}{\left|{x(n)}\right|^{2}}x(n)\label{xIMDn}
\end{equation}
The filter coefficients ${{\textbf{h}}_{{\rm {IMD}}}^{\textmd{o}}}={[{h_{{\rm {IMD}},1}^{\textmd{o}}},{h_{{\rm {IMD}},2}^{\textmd{o}}},\ldots,{h_{{\rm {IMD}},N}^{\textmd{o}}}]^{T}}$ and ${{\textbf{g}}_{{\rm {IMD}}}^{\textmd{o}}}={[{g_{{\rm {IMD}},1}^{\textmd{o}}},{g_{{\rm {IMD}},2}^{\textmd{o}}},\ldots,{g_{{\rm {IMD}},N}^{\textmd{o}}}]^{T}}$, where $N<M$, respectively represent the end-to-end channel impulse responses of the IMD SI component ${{\textbf{x}}_{{\rm {IMD}}}}(n)$ and its complex conjugate ${{\textbf{x}}_{{\rm {IMD}}}}^{*}(n)$. In \eqref{un_vec}, the interference term $v(n)$ represents the thermal noise during the digital SI cancellation process, which is assumed to be a zero-mean complex-valued additive white Gaussian noise (AWGN), and its variance $\sigma_{v}^{2}$ is determined by \cite{Korpi2014,Korpi2014a}	
\begin{eqnarray}
\sigma_{v}^{2}\! = \!\frac{{k_{{\rm {BB}}}{k_{{\rm {LNA}}}}}k_{{\rm {RIQ}}}{p_{{\rm {sen}}}}}{{\rm {SNR}_{{\rm {req}}}}}\label{sigmav}
\end{eqnarray}
The quantization noise $q(n)$ in \eqref{un_vec} is assumed to be another zero-mean AWGN process, whose variance $\sigma_{q}^{2}$ is computed as \cite{Gu_Book}
\begin{equation}
\sigma_{q}^{2}\! = \! \frac{p_{{\rm ADC}}}{10^{6.02\beta+4.76-{\rm PAPR}/10}}\label{sigmaq2}
\end{equation}
Physical meanings of other unmentioned parameters used from \eqref{xIMDn} to \eqref{sigmaq2} are provided in Table \ref{denotation}.
%
%

From \eqref{dn_vec}, it is clear that in a feasible FD transceiver, the baseband signal before digital SI cancellation $d(n)$ is composed of various interference components, including SI ${\textbf{x}}(n)$, IMD SI ${\textbf{x}}_{{\rm {IMD}}}(n)$, thermal noise $v(n)$ and their image counterparts, as well as quantization noise $q(n)$. In order to ascertain which components can be counted as primary interference under certain circumstances, simulations were carried out to illustrate their power relations within FD transceivers. According to the suggestions in \cite{Korpi2014}, two types of practical FD transceivers were considered, the system parameters of which are listed in Table \ref{ParaFDDCT}. The main difference lies in their different analog SI cancellation capabilities. The RF separation and attenuation capabilities of the Type 1 transceiver are 40 dB and 30 dB respectively, while in a Type 2 transceiver, both were 10 dB lower. Consequently, the Type 2 transceiver exhibited an inferior analog SI cancellation capability compared to its Type 1 counterpart, weakening both the received signal of interest $x_{{\rm {SOI}}}(n)$ and the thermal noise $v(n)$.

\begin{figure}[t!]
	\centering \subfigure[Type 1 FD transceiver]{ \label{ComponentAnalysis:4030} 
		\begin{minipage}[b]{0.48\textwidth}%
			\centering \includegraphics[width=1\textwidth]{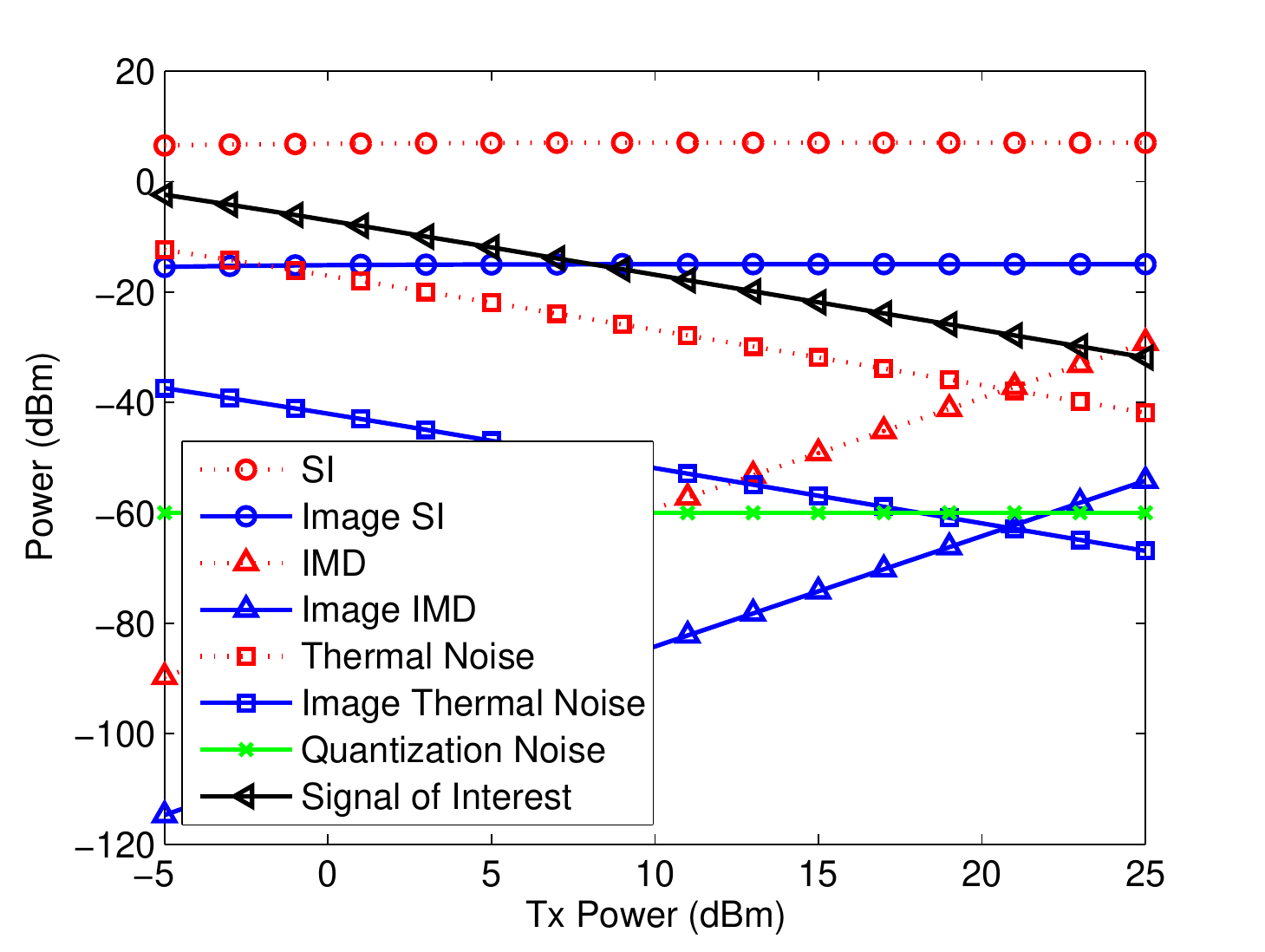} %
	\end{minipage}}\hfill{}\subfigure[Type 2 FD transceiver]{ \label{ComponentAnalysis:3020} 
		\begin{minipage}[b]{0.48\textwidth}%
			\centering \includegraphics[width=1\textwidth]{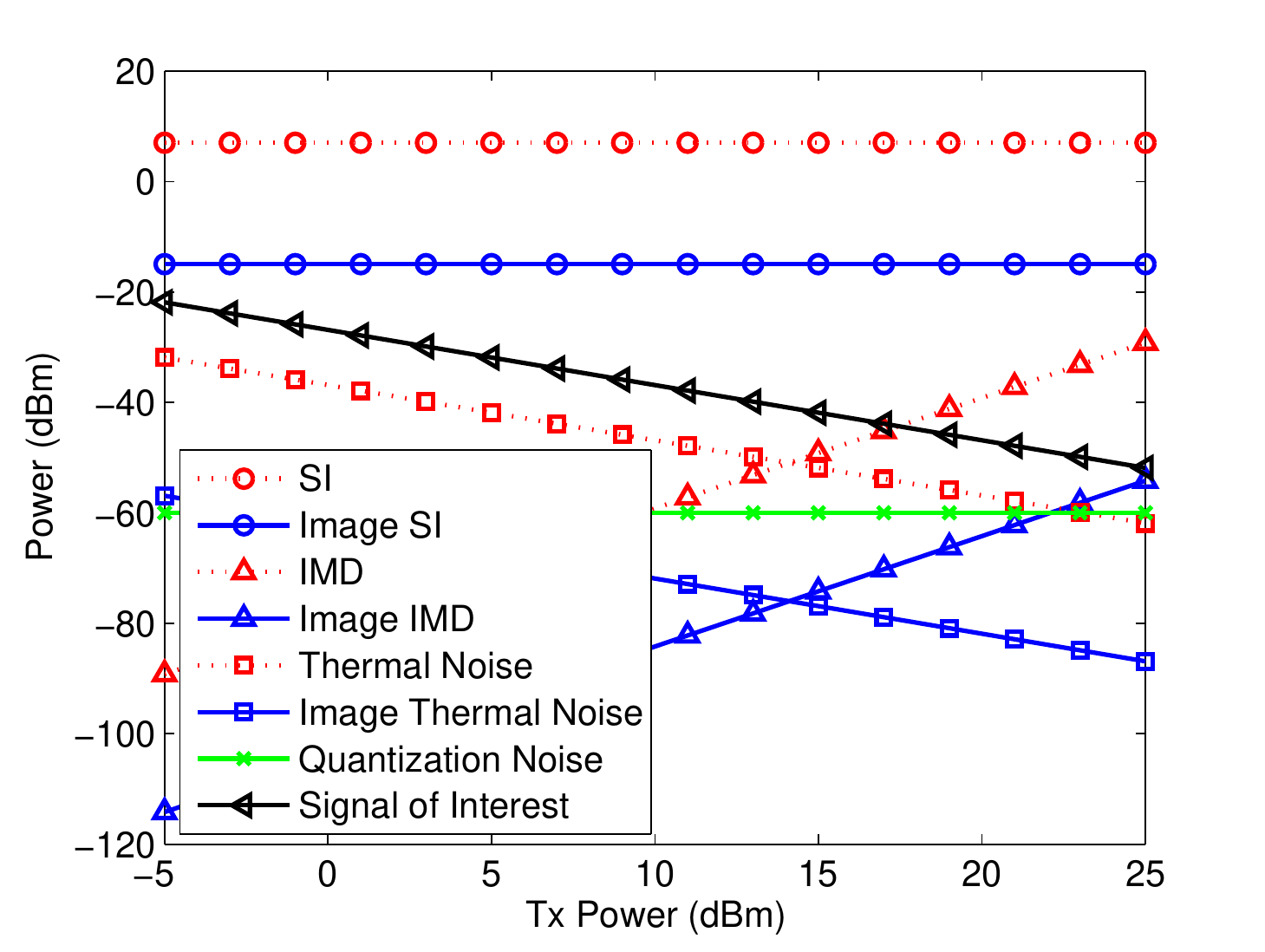} %
	\end{minipage}} \caption{Power comparison among different signal components in representative
		FD transceivers, before digital SI cancellation and against different levels
		of transmit powers. (a) Type 1 FD transceiver, and (b) Type 2 FD transceiver.}
	\label{ComponentAnalysis} 
\end{figure}

As shown in Fig. \ref{ComponentAnalysis:4030}, in a Type 1 FD transceiver, the SI component $\textbf{x}(n)$ and its image counterpart $\textbf{x}^{*}(n)$  are both the dominant interference to the signal of interest in the entire transmit power range, and when the transmit power becomes higher, the IMD SI component ${\textbf{x}}_{{\rm {IMD}}}(n)$ linearly increased to become another major interference \cite{Korpi2014}. On the other hand, in Fig. \ref{ComponentAnalysis:3020}, when the transmit power of a Type 2 FD transceiver went above 20 dBm, the thermal noise $v(n)$ became weaker than the quantization noise $q(n)$ and the image IMD SI ${\textbf{x}}_{{\rm {IMD}}}^{*}(n)$. This is because either a stronger nonlinear SI ${\textbf{x}}_{{\rm {IMD}}}(n)$ or a less efficient analog cancellation results in a lower receiver variable gain amplifier (VGA) gain $k_{{\rm {BB}}}$.
\begin{table}[t!]
	\scriptsize
	\centering \caption{Used notations and symbols.}
	\label{denotation} %
	\begin{tabular}{|c|c|}
		\hline
		Symbol & Denotation\tabularnewline
		\hline
		$k_{{\rm {VGA}}}$ & Tx variable gain amplifier (VGA) gain\tabularnewline
		\hline
		$k_{{\rm {BB}}}$ & Rx VGA gain\tabularnewline

		\hline
		$k_{{\rm {LNA}}}$ & LNA gain\tabularnewline
		\hline
		$\alpha_{0}$ & PA gain\tabularnewline
		\hline
		$\alpha_{1}$ & Gain of PA nonlinearity \tabularnewline
		\hline
		$k_{{\rm {TIQ}}}$ & Tx IQ mixer gain\tabularnewline
		\hline
		$k_{{\rm {RIQ}}}$ & Rx IQ mixer gain\tabularnewline
		\hline
		$f_{{\rm {PA}}}(n)$ & Memory polynomials of PA\tabularnewline
		\hline
		${f_{{\rm {RFE}}}}(n)$ & Estimation error of analog cancellation\tabularnewline
		\hline
		${p_{{\rm {sen}}}}$ & Receiver sensitivity\tabularnewline
		\hline
		${\rm {SNR}_{{\rm {req}}}}$ & SNR requirement\tabularnewline
		\hline
		${p_{{\rm {ADC}}}}$ & Dynamic range of ADC\tabularnewline
		\hline
		${\rm {PAPR}}$ & Peak-to-average-power ratio \tabularnewline
		\hline
		$\beta$ &  ADC bits \tabularnewline
		\hline
	\end{tabular}
\end{table}
\begin{table}[t!]
	\centering \scriptsize \caption{System parameters of typical FD transceivers.}
	\begin{tabular}{|c||c|}
		\hline
		Parameter & Value  \tabularnewline
		\hline
		Receiver sensitivity & -89 dBm   \tabularnewline
		\hline
		SNR requirement & 15 dB  \tabularnewline
		\hline
		Thermal noise floor & -104 dBm   \tabularnewline
		\hline
		RF separation  & %
		\begin{tabular}{@{}l@{}}
			40 dB (Type 1) \tabularnewline
			30 dB (Type 2) \tabularnewline
		\end{tabular} \tabularnewline
		\cline{1-2}
		RF attenuation  & %
		\begin{tabular}{@{}l@{}}
			30 dB (Type 1) \tabularnewline
			20 dB (Type 2) \tabularnewline
		\end{tabular}  \tabularnewline
		\hline
		IRR & 25 dB   \tabularnewline
		\hline
		Tx mixer gain & 6 dB   \tabularnewline
		\hline
		Rx mixer gain & 6 dB   \tabularnewline
		\hline
		PA gain & 27 dB  \tabularnewline
		\hline
		PA IIP3 & 20 dBm   \tabularnewline
		\hline
		LNA gain & 25 dB  \tabularnewline
		\hline
		Transmit power & ${\rm {-5\sim25}}$ dBm   \tabularnewline
		\hline
		ADC dynamic range & 7 dB  \tabularnewline
		\hline
		ADC bits  & 12  \tabularnewline
		\hline
		Peak-to-average-power ratio & 10 dB   \tabularnewline
		\hline
	\end{tabular}\label{ParaFDDCT}
\end{table}
\section{Conventional Augmented LMS based SI Canceller and Its Performance Analysis}\label{pfmAna}
In this section, we provide a comprehensive mean and mean square convergence analysis of the conventional  augmented LMS, employed in \cite{Sakai2016a} as a DSP-assisted analog SI cancellation process, in the presence of both frequency-dependent Tx and Rx I/Q imbalance and PA nonlinear distortion. For rigor, the proposed analysis covers both the cases of low and high transmit powers, as discussed in Section \ref{model}.

For the compactness of analysis, we first represent the widely linear model in \eqref{dn_vec} in an augmented form, given by
\begin{equation}
d(n)=\textbf{x}^{aT}(n)\textbf{w}^{a\textmd{o}}+u(n)\label{dn_aug}
\end{equation}
where $\textbf{x}^{a}(n)=[\textbf{x}^{T}(n),\textbf{x}^{H}(n)]^{T}$
is the $2M\times1$ augmented SI vector. In real-world wireless communications, the SI waveform $x(n)$ is always oversampled and bandlimited, and consequently non-white. However, for the simplicity of the analysis, we here assume $x(n)$ is critically sampled, so that, it can be modeled as a zero-mean proper white Gaussian random variable with variance $\sigma_{x}^{2}$. Note that the Gaussianity and properness assumptions on $x(n)$ are valid for wideband OFDM waveforms. Indeed, the work in \cite{Wei2010} has verified that a bandlimited uncoded OFDM symbol converges to a proper Gaussian random process as the number of subcarriers increases. The vector $\textbf{w}^{a\textmd{o}}=[\textbf{h}^{\textmd{o}T},\textbf{g}^{\textmd{o}T}]^{T}$
contains the augmented end-to-end system impulse responses, which models the transmit
and receive frequency-dependent I/Q imbalance, PA distortion, and the residual of analog SI cancellation.

The augmented LMS estimates the set of system parameters $\textbf{w}^{a\textmd{o}}$ by minimizing the MSE cost function $J^{a}(n)$, defined as \cite{Javidi2008,Mandic_Book2,Xia2012,Sakai2016a}
\begin{equation}
{J^{a}}(n)=E[\left|e^{a}(n)\right|^{2}]=E[e^{a}(n)e^{a*}(n)]\label{MSEDef}
\end{equation}
where $e^{a}(n)$ is the instantaneous output error or residual SI, given by
\begin{equation}
e^{a}(n)=d(n)-\textbf{x}^{aT}(n)\textbf{w}^{a}(n)\label{en}
\end{equation}
and the augmented weight vector, that is, ${\textbf{w}}^{a}(n)$, is updated as
\begin{equation}
\textbf{w}^{a}(n+1)=\textbf{w}^{a}(n)+\mu e^{a}(n)\textbf{x}^{a*}(n)\label{wnIter}
\end{equation}
where $\mu$ is the step size.
\vspace{-0.3cm}
\subsection{Mean Convergence Analysis}
\label{meanConvAna}
Upon introducing the $2M\times1$ weight error vector
\begin{equation}
\widetilde{\textbf{w}}^{a}(n)=\textbf{w}^{a}(n)-\textbf{w}^{a\textmd{o}}\label{misalign}
\end{equation}
the output residual SI $e^{a}(n)$ in \eqref{en} now becomes
\begin{equation}
e^{a}(n)=u(n)-\textbf{x}^{aT}(n)\widetilde{\textbf{w}}^{a}(n)\label{en2}
\end{equation}
From \eqref{wnIter}, the recursion for the update of the weight error
vector ${\widetilde{\textbf{w}}}^{a}(n)$ can be derived as
\begin{equation}
\widetilde{\textbf{w}}^{a}(n\!+\!1)\!=\![\textbf{I}_{2M}\!-\!\mu\textbf{x}^{a*}(n)\textbf{x}^{aT}(n)]\widetilde{\textbf{w}}^{a}(n)\!+\!\mu u(n)\textbf{x}^{a*}(n)\label{misalgnIter}
\end{equation}
%
By applying the statistical expectation operator $E[\cdot]$ to both sides of \eqref{misalgnIter} and upon employing the standard independence assumptions \cite{Fisher_1983,Douglas_2010}, that is, the composite noise $u(n)$ is statistically independent of any other variable within the augmented LMS, and ${\widetilde{\textbf{w}}}^{a}(n)$ is statistically independent of the augmented SI input vector $\textbf{x}^{a}(n)$, we have
\begin{equation}
E[\widetilde{\textbf{w}}^{a}(n\!+\!1)]\!=\!(\textbf{I}_{2M}\!-\!\mu\textbf{R}^{a*})E[\widetilde{\textbf{w}}^{a}(n)]\!+\!\mu E[u(n)\textbf{x}^{a*}(n)]\label{misalgnIterExp}
\end{equation}
where ${\textbf{R}}^{a}$ is the covariance matrix of the augmented SI vector $\textbf{x}^{a}(n)$, defined as
\begin{equation}
\textbf{R}^{a}=E[\textbf{x}^{a}(n)\textbf{x}^{aH}(n)]=\sigma_{x}^{2}\textbf{I}_{2M}
\end{equation}
Therefore, the convergence of  the augmented LMS in the mean weight error sense is guaranteed if the
step-size $\mu$ satisfies \cite{Douglas_2010}
\begin{equation}
0<\mu<\frac{2}{\lambda_{\max}[\textbf{R}^{a}]}=\frac{2}{\sigma_{x}^{2}}\label{stepSizeMeanCvg}
\end{equation}
From Fig. \ref{ComponentAnalysis:4030} and Fig. \ref{ComponentAnalysis:3020}
in Section \ref{model}, we observe that the relative power relationships
among different SI components vary as the transmit power changes.
Therefore, in order to accurately describe the statistical mean behavior
of the  augmented LMS based SI canceller, we consider the following two case studies.

%
\textit{1) Low transmit power.} In this situation, both the nonlinear distortion component $\textbf{x}_{{\rm {IMD}}}(n)$ and the quantization noise $q(n)$ at the receiver are negligible, since their powers are much weaker than that of the thermal noise $v(n)$ in \eqref{un_vec}. Therefore, from \eqref{dn_aug}, we have $u(n)\simeq v(n)$, and hence, at the steady-state, as, $n\rightarrow\infty$, from \eqref{misalgnIterExp} we have
\begin{equation}
E[{\widetilde{\textbf{w}}}^{a}(\infty)]={\bf {0}}\label{wevInfty0}
\end{equation}
\vspace{-0.3cm}

\textit{2) High transmit power.} When the transmit power is high, the PA nonlinear distortion component $\textbf{x}_{{\rm {IMD}}}(n)$ becomes the third strongest interference among all the imperfections considered. After employing the independence assumptions and \eqref{un_vec}, the second term on the right hand side (RHS) of \eqref{misalgnIterExp} can be derived as 
\begin{align}
\!\!&E[u(n){\textbf{x}}^{a*}(n)]\!=\!E[\big({\textbf{x}}_{{\rm {IMD}}}^{T}(n){{\textbf{h}}_{{\rm {IMD}}}^{\textmd{o}}}\!+\!{\textbf{x}}_{{\rm {IMD}}}^{H}(n){{\textbf{g}}_{{\rm {IMD}}}^{\textmd{o}}}\big){\textbf{x}}^{a*}(n)]\nonumber \\
\!\! & =k_{{\rm {TIQ}}}^{3/2}\Big\{\sum_{i=1}^{N}E[{{h_{{\rm {IMD}},i}}{{\left|{x(n-i+1)}\right|}^{2}}x(n-i+1)}{\textbf{x}}^{a*}(n)]\nonumber \\
\!\! & +\sum_{i=1}^{N}E[{{g_{{\rm {IMD}},i}}{{\left|{x(n-i+1)}\right|}^{2}}x^{*}(n-i+1)}{\textbf{x}}^{a*}(n)]\Big\}\nonumber \\
\!\! & =k_{{\rm {TIQ}}}^{3/2}E[{\left|{x(n)}\right|^{4}}]{[{{\textbf{h}}_{{\rm {IMD}}}^{{\textmd{o}}T}}, {\textbf{0}}_{M-N}^{T}, {{\textbf{g}}_{{\rm {IMD}}}^{{\textmd{o}}T}}, {\textbf{0}}_{M-N}^{T}]^{T}}\nonumber \\
\!\! & =2k_{{\rm {TIQ}}}^{3/2}\sigma_{x}^{4}{[{{\textbf{h}}_{{\rm {IMD}}}^{{\textmd{o}}T}}, {\textbf{0}}_{M-N}^{T}, {{\textbf{g}}_{{\rm {IMD}}}^{{\textmd{o}}T}}, {\textbf{0}}_{M-N}^{T}]^{T}}\label{Eunxaconjn}
\end{align}
The last step is performed with the help of the Gaussian fourth order moment factorizing theorem \cite{Mandic2015,Xia2017,Xia2017b}, and since $x(n)$ is proper (second order circular), we have $E[{\left|{x(n)}\right|^{4}}]=2\sigma_{x}^{4}$. From \eqref{misalgnIterExp}, the steady-state
value of the weight error vector, that is, $E[{\widetilde{\textbf{w}}}^{a}(\infty)]$,
 can be evaluated as
\begin{eqnarray}
E[{\widetilde{\textbf{w}}}^{a}(\infty)]\!\!\!\! & = & \!\!\!\!{\left({{\textbf{R}}^{a*}}\right)^{-1}}E[u(n){\textbf{x}}^{a*}(n)]\nonumber \\
\!\!\!\! & = & \!\!\!\!2k_{{\rm {TIQ}}}^{3/2}\sigma_{x}^{2}{[{{\textbf{h}}_{{\rm {IMD}}}^{{\textmd{o}}T}}, {\textbf{0}}_{M-N}^{T}, {{\textbf{g}}_{{\rm {IMD}}}^{{\textmd{o}}T}}, {\textbf{0}}_{M-N}^{T}]^{T}}\label{wevInfty}
\end{eqnarray}

\textit{Remark 1:} The upper bound on the step size $\mu$ for the
mean convergence of  augmented LMS for a low transmit power FD transceiver is identical
to that for a high transmit power one. At the steady state, when the
transmit power is low, the augmented LMS converges in the mean to the optimal
weight coefficients associated with ${\textbf{x}}^{a}(n)$, that is, $\textbf{w}^{a\textmd{o}}$ in \eqref{dn_aug}, in an
unbiased manner. However, as indicated by \eqref{wevInfty}, when
the transmit power is high, this yields a bias in the estimation of
$2N$ out of $2M$ entries of the weight error vector $E[{\widetilde{\textbf{w}}}^{a}(\infty)]$,
quantified by $2k_{{\rm {TIQ}}}^{3/2}\sigma_{x}^{2}{[{{\textbf{h}}_{{\rm {IMD}}}^{{\textmd{o}}T}}, {{\textbf{g}}_{{\rm {IMD}}}^{{\textmd{o}}T}}]^{T}}$.
The level of this bias depends upon the level of undermodeling, that
is, the transmitter mixer gain $k_{{\rm {TIQ}}}$, the transmit SI signal
power $\sigma_{x}^{2}$, and the channel impulse responses associated with the IMD
SI components, that is, ${{\textbf{h}}_{{\rm {IMD}}}^{{\textmd{o}}}}$
and ${{\textbf{g}}_{{\rm {IMD}}}^{{\textmd{o}}}}$.
%
%
\subsection{Mean Square Convergence Analysis}
From \eqref{MSEDef} and \eqref{en2}, and again by employing the
standard independence assumptions stated in Section \ref{meanConvAna},
the MSE of  augmented LMS based SI canceller, that is, $J^{a}(n)$, can be
further evaluated as
\begin{align}
\!{J^{a}}(n)\! & =  \!E[{{\widetilde{\textbf{w}}}^{aH}}(n){{\textbf{x}}^{a}}^{*}(n){\textbf{x}}^{aT}(n){{\widetilde{\textbf{w}}}^{a}}(n)]\!+\!E[{\left|{u(n)}\right|^{2}}]\nonumber \\
\! & -  \!E[{u^{*}}(n){{\textbf{x}}^{aT}}(n){\widetilde{\textbf{w}}}^{a}(n)]\!-\!E[{u(n){\widetilde{\textbf{w}}}^{aH}(n)}{\textbf{x}}^{a*}(n)]\nonumber \\
\! & =  \!\!{\rm {Tr}}[\textbf{R}^{a*}\textbf{K}^{a}(n)]\!\!+\!\!E[{\left|{u(n)}\right|^{2}}]\!-\!E[{u^{*}}(n){{\textbf{x}}^{aT}}(n)]E[{\widetilde{\textbf{w}}}^{a}(n)]\nonumber \\
\! & -  \!E[{u(n){\textbf{x}}^{aH}(n)]E[{\widetilde{\textbf{w}}}^{a*}(n)}]\label{MSE}
\end{align}
where ${{\textbf{K}}^{a}}(n)=E[{\widetilde{\textbf{w}}}^{a}(n){{\widetilde{\textbf{w}}}^{aH}}(n)]$
is the covariance matrix of the augmented weight error vector ${\widetilde{\textbf{w}}}^{a}(n)$ \cite{Mandic2015,Xia2017,Xia2017b}.
It can be observed from \eqref{MSE} that the mean square convergence
analysis of  augmented LMS now rests upon both the first and second
order statistical properties of ${\widetilde{\textbf{w}}}^{a}(n)$.
To this end, we first apply the Hermitian operator $(\cdot)^{H}$
to both sides of \eqref{misalgnIter}, to yield
\begin{equation}
{\widetilde{\textbf{w}}}^{aH}(n\!+\!1)\!=\!{\widetilde{\textbf{w}}}^{aH}(n)[{\textbf{I}}_{2M}\!-\!\mu{\textbf{x}}^{a*}(n){\textbf{x}}^{aT}(n)]\!+\!\mu u^{*}(n){\textbf{x}}^{aT}(n)\label{misalgnIterH}
\end{equation}
Upon post-multiplying both sides of \eqref{misalgnIter} by ${\widetilde{\textbf{w}}}^{aH}(n+1)$ in \eqref{misalgnIterH} and taking the statistical expectation $E[\cdot]$, the evolution of the weight error covariance matrix ${{\textbf{K}}^{a}}(n)$ now becomes
\begin{eqnarray}
\!\!{\textbf{K}}^{a}(n\!+\!1)\!\!\!\! & = & \!\!\!\!(1\!-\!2\mu\sigma_{x}^{2}){\textbf{K}}^{a}(n)\!+\!2\mu\Re[{{\textbf{Q}}_{3}}(n)]\nonumber \\
\!\!\!\! & + & \!\!\!\!{\mu^{2}}[{{\textbf{Q}}_{1}}(n)\!+\!{{{\textbf{Q}}_{2}}(n)}\!-\!{{{\textbf{Q}}_{4}}(n)}\!-\!{{{\textbf{Q}}_{5}}(n)}]\label{evoofKn}
\end{eqnarray}
where
\begin{eqnarray}
{{{\textbf{Q}}_{1}}(n)}\!\!\!\! & = & \!\!\!\!E[{{{\left|{u(n)}\right|}^{2}}{{\textbf{x}}^{a*}}(n){\textbf{x}}^{aT}(n)}]\nonumber \\
{{{\textbf{Q}}_{2}}(n)}\!\!\!\! & = & \!\!\!\!E[{{\textbf{x}}^{a*}}(n){\textbf{x}}^{aT}(n){\widetilde{\textbf{w}}}^{a}(n){{\widetilde{\textbf{w}}}^{aH}}(n){{\textbf{x}}^{a*}}(n){\textbf{x}}^{aT}(n)]\nonumber \\
{{{\textbf{Q}}_{3}}(n)}\!\!\!\! & = & \!\!\!\!E[{u(n){\textbf{x}}^{a*}(n)]E[{\widetilde{\textbf{w}}}^{aH}(n)}]\nonumber \\
{{{\textbf{Q}}_{4}}(n)}\!\!\!\! & = & \!\!\!\!E[{u(n){{\textbf{x}}^{a*}}(n){\textbf{x}}^{aH}(n){\textbf{x}}^{a*}(n)}]E[{{\widetilde{\textbf{w}}}^{aT}}(n)]\nonumber \\
{{{\textbf{Q}}_{5}}(n)}\!\!\!\! & = & \!\!\!\!E[{{u^{*}}(n){{\textbf{x}}^{a*}}(n){{\textbf{x}}^{aT}}(n){\textbf{x}}^{a}(n)}]E[{\widetilde{\textbf{w}}}^{a}(n)]\label{Qs}
\end{eqnarray}
It can be observed that the term ${{{\textbf{Q}}_{2}}(n)}$ is independent of the IMD component ${\textbf{x}}_{{\rm {IMD}}}(n)$, and hence
\begin{equation}
{{\textbf{Q}}_{2}}(n)\!=\!\sigma_{x}^{4}{{\textbf{K}}^{a}}(n)\!+\!{{\textbf{P}}^{a*}}{{{\textbf{K}}^{aT}}(n)}{\textbf{P}}^{a}\!+\!2M\sigma_{x}^{4}{{\textbf{I}}_{2M}}{\rm {Tr}}[{{\textbf{K}}^{a}}(n)]\!\!\label{Expection2}
\end{equation}
where ${{\textbf{P}}^{a}}$ is the pseudocovariance matrix of the augmented SI vector ${{\textbf{x}}^{a}}(n)$ \cite{Xia2017,Xia2017b,Xia2018a}, given by
\[
{{\textbf{P}}^{a}}=E[{{\textbf{x}}^{a}}(n){\textbf{x}}^{aT}(n)]=\left[{\begin{array}{cc}
{\textbf{0}} & {\sigma_{x}^{2}{{\textbf{I}}_{M}}}\\
{\sigma_{x}^{2}{{\textbf{I}}_{M}}} & {\textbf{0}}
\end{array}}\right]
\]
%
The term ${\rm {Tr}}[{{\textbf{K}}^{a}}(n)]$ on the RHS of \eqref{Expection2} can be decomposed as ${\textbf{1}}_{2M}^{T}{\bm{\kappa}}^{a}(n)$, where ${\bm{\kappa}}^{a}(n)$ is a $2M\times1$ vector, whose entries are the diagonal elements of ${{\textbf{K}}^{a}}(n)$, defined as
\begin{eqnarray}
\bm{\kappa}^{a}(n)=\big[E[|\widetilde{w}_{1}^{a}(n)|^{2}],E[|\widetilde{w}_{2}^{a}(n)|^{2}],\ldots,E[|\widetilde{w}_{2M}^{a}(n)|^{2}]\big]^{T}\label{Eq:vec_kappa}
\end{eqnarray}
Then, based on \eqref{evoofKn}, the evolution of ${\bm{\kappa}}^{a}(n)$ becomes
\begin{align}
\!\!&\!{\bm{\kappa}}^{a}(n\!+\!1)\nonumber\\
\!\!\!\!&\!=\!\underbrace{\{{{\textbf{I}}_{2M}}\!-\!2\mu\sigma_{x}^{2}{{\textbf{I}}_{2M}}\!+\!2{\mu^{2}}\sigma_{x}^{4}{{\textbf{I}}_{2M}}\!+\!{\mu^{2}}\sigma_{x}^{4}{{\textbf{1}}_{2M}}{\textbf{1}}_{2M}^{T}\}}_{{\textbf{F}}^{a}}{\bm{\kappa}}^{a}(n)\nonumber\\
\!\!\!\!&\!+\!2\mu{\rm {diag}}\{\Re[{{{\textbf{Q}}_{3}}(n)}]\}\!+\!{\mu^{2}}{\rm {diag}}\{{{\textbf{Q}}_{1}}(n)\!-\!{{{\textbf{Q}}_{4}}(n)}\!-\!{{{\textbf{Q}}_{5}}(n)}\!\}\! \label{evoofkn}
\end{align}
The convergence of the recursion for the vector ${\bm{\kappa}}^{a}(n)$
in \eqref{evoofkn} is subject to two conditions: 1) the terms $\textbf{Q}_{1}(n)$,
${{\textbf{Q}}_{3}}(n)$, ${{\textbf{Q}}_{4}}(n)$ and ${{\textbf{Q}}_{5}}(n)$
are bounded, which is guaranteed if $E[{\widetilde{\textbf{w}}}^{a}(n)]$
is bounded; 2) all the eigenvalues of the transition matrix ${\bf {F}}^{a}$
are less than unity \cite{Mayyas2005,Xia2018}. From \eqref{stepSizeMeanCvg},
condition 1) holds when $0<\mu<2/\sigma_{x}^{2}$. Furthermore, the
eigenvalues of ${{\textbf{F}}^{a}}$, denoted by $\lambda_{i}^{a}$,
where $i=1,2,\ldots,2M$, can be obtained by solving $\det[{{\textbf{F}}^{a}}-\lambda_{i}^{a}{{\textbf{I}}_{2M}}]=0$,
while from \eqref{evoofkn}, it is easy to find that ${{\textbf{F}}^{a}}$
is Toeplitz, for which the diagonal elements are $1-2\mu+3\mu^{2}\sigma_{x}^{4}$,
and off-diagonal ones are $\mu^{2}\sigma_{x}^{4}$. Hence, after some algebraic manipulations, we have
\begin{eqnarray*}
\lambda_{1}^{a}\!\!\!\!&=&\!\!\!\! 1-{\mu}\sigma_{x}^{2}+(2M+2){\mu}^{2}\sigma_{x}^{4}\\
\lambda_{j}^{a}\!\!\!\!&=&\!\!\!\!1-{\mu}\sigma_{x}^{2}+2{\mu}^{2}\sigma_{x}^{4},~j=2,3,\ldots,2M
\end{eqnarray*}
Note that since $M\geq1$, we have $\lambda_{1}^{a}>\lambda_{j}^{a}$ for $j=2,3,\ldots,2M$, and hence, condition 2) is satisfied if $\lambda_{1}^{a}<1$, to yield
\begin{equation}
0<\mu<\frac{1}{(M+1)\sigma_{x}^{2}}\label{stepSizeMeanSqCvg}
\end{equation}

\textit{Remark 2:} The upper bound in \eqref{stepSizeMeanSqCvg} is
tighter than that in Condition 1), and therefore, the mean square convergence of  the augmented LMS based SI canceller in the presence of frequency-dependent
IQ imbalance and PA distortion is guaranteed if the step-size $\mu$ satisfies \eqref{stepSizeMeanSqCvg}.
\subsection{Steady State Analysis}
Suppose that step-size $\mu$ is chosen such that the mean square stability of  augmented LMS is guaranteed. Consider $n\rightarrow\infty$, then based on \eqref{MSE} and \eqref{Eq:vec_kappa}, the steady-state MSE $J^{a}(\infty)$ can be expressed as
\begin{equation}
{J^{a}}(\infty)\!=\!\sigma_{x}^{2}{\textbf{1}}_{2M}^{T}{\bm{\kappa}}^{a}(\infty)\!+\!E[{\left|{u(\infty)}\right|^{2}}]\!-\!2{\rm {Tr}}(\Re[{{{\textbf{Q}}_{3}}(\infty)}])\label{ACLMS_SS_MSE}
\end{equation}
where, based on \eqref{evoofkn}, ${\bm{\kappa}}^{a}(\infty)$ can be derived as
\begin{align}
{\bm{\kappa}}^{a}(\infty) & =2\mu({{\textbf{I}}_{2M}}-{{\textbf{F}}^{a}})^{-1}\big[{\rm {diag}}\{\Re[{{{\textbf{Q}}_{3}}(\infty)}]\}\nonumber \\
 & +\!{\mu}{\rm {diag}}\{{{\textbf{Q}}_{1}}(\infty)\!-\!{{{\textbf{Q}}_{4}}(\infty)}\!-\!{{{\textbf{Q}}_{5}}(\infty)}\}\big]\label{evoofkinfty}
\end{align}
Similar to the mean convergence analysis in Section \ref{meanConvAna}, in order to evaluate the terms in \eqref{ACLMS_SS_MSE} and \eqref{evoofkinfty}, which involves the overall noise $u(\infty)$, we need to consider the levels of the transmit power.

\textit{1) Low transmit power.} In this case, the power of the thermal noise $v(\infty)$ in $u(\infty)$ is much higher than  the power of the nonlinear distortion component $\textbf{x}_{{\rm {IMD}}}(\infty)$
and the quantization noise $q(\infty)$, so that, $u(n)\simeq v(n)$, and $E[{\left|{u(\infty)}\right|^{2}}]\simeq\sigma_{v}^{2}$. Upon inserting \eqref{wevInfty0} into \eqref{Qs}, we obtain ${{{\textbf{Q}}_{1}}(\infty)}=\sigma_{v}^{2}\sigma_{x}^{2}{{\textbf{I}}_{2M}}$, and ${{{\textbf{Q}}_{3}}(\infty)}={{{\textbf{Q}}_{4}}(\infty)}={{{\textbf{Q}}_{5}}(\infty)}={\textbf{0}}$, therefore, ${{\bm{\kappa}}^{a}}(\infty)$ in \eqref{evoofkinfty} can be simplified as
\begin{equation*}
{{\bm{\kappa}}^{a}}(\infty)=\frac{{\mu\sigma_{v}^{2}}}{{2(1-\mu(M+1)\sigma_{x}^{2})}}{{\textbf{1}}_{2M}}\label{kappaInftySmall}
\end{equation*}
Upon substituting into \eqref{ACLMS_SS_MSE}, this yields
\begin{equation}
J_{{\rm {low}}}^{a}(\infty)\!=\!\frac{{(1-\mu\sigma_{x}^{2})\sigma_{v}^{2}}}{{1-\mu(M+1)\sigma_{x}^{2}}}\label{MSElowTx}
\end{equation}
The achievable SINR is defined as a relative power ratio between the received signal of interest $x_{{\rm {SOI}}}(n)$ and the residual SI $e^{a}(n)$. According to the analysis in \cite{Korpi2014}, the power of $x_{{\rm {SOI}}}(n)$ can be evaluated as
\begin{equation}
p_{x_{{\rm {SOI}}}}={p_{{\rm {sen}}}}{k_{{\rm {LNA}}}}{k_{{\rm {BB}}}}{k_{{\rm {RIQ}}}}\label{pxSOI}
\end{equation}
where the Rx VGA gain ${k_{{\rm {BB}}}}$ ensures the received signal fit within the voltage range of the ADC, and can be calculated as \cite{Korpi2014a}
\begin{align*}\label{kBB}
\!{k_{\rm BB}} &\!=\!\frac{p_{\rm ADC}}{{k_{\rm LNA}}{k_{\rm RIQ}}} \nonumber\\
&\cdot \frac{1}{ [\alpha_{0}^{2}k_{{\rm {VGA}}}k_{{\rm {TIQ}}}\sigma_{x}^{2}
	\!+\!\alpha_{1}^{2}k_{{\rm {VGA}}}^3k_{{\rm {TIQ}}}^{3}\sigma_{x}^{6}]{\left\Vert {f_{{\rm {RFE}}}}(n)\right\Vert ^{2}}\negthinspace+\negthinspace{p_{\rm sen}}}
\end{align*}
From \eqref{sigmav}, \eqref{MSElowTx} and \eqref{pxSOI}, the achievable SINR of  augmented LMS at the steady state can be evaluated as
\begin{equation}
{{\rm {SINR}}}_{{\rm {low}}}\!=\!\frac{{p_{{\rm {x}}_{{\rm {SOI}}}}}}{J_{{\rm {low}}}^{a}(\infty)}\!=\!\frac{{1-\mu(M+1)\sigma_{x}^{2}}}{{1-\mu\sigma_{x}^{2}}}{\rm {SNR}}{_{{\rm {req}}}}\label{SINRAchv}
\end{equation}

\textit{2) High transmit power.} In this case, with typical parameters of FD transceivers given in Table \ref{ParaFDDCT},
the power of the SI component $\sigma_{x}^{2}$ is guaranteed to be
less than unity. Therefore, in \eqref{evoofkinfty}, the Euclidean
norm of ${\mu}{\rm {diag}}\{{{\textbf{Q}}_{1}}(\infty)-{{{\textbf{Q}}_{4}}(\infty)}-{{{\textbf{Q}}_{5}}(\infty)}\}$
is much smaller than that of ${\rm {diag}}\{\Re[{{{\textbf{Q}}_{3}}(\infty)}]\}$,
as the terms ${{\textbf{Q}}_{1}}(\infty)$, ${{{\textbf{Q}}_{4}}(\infty)}$
and ${{{\textbf{Q}}_{5}}(\infty)}$ contain a larger amount of higher order
SI components. This situation is more pronounced when a smaller step-size
$\mu$ is chosen, and based on the analysis in Appendix \ref{App:AppendixA},
we have
\begin{align}
 & {{\mu}{\rm {diag}}\{{{\textbf{Q}}_{1}}(\infty)\!-\!{{{\textbf{Q}}_{4}}(\infty)}\!-\!{{{\textbf{Q}}_{5}}(\infty)}\}\!+\!{\rm {diag}}}\{\Re[{{{\textbf{Q}}_{3}}(\infty)}]\}\nonumber \\
 & \!=\!\mu\big(E[{\left|{v(\infty)}\right|^{2}}]\!+\!E{\left|{q(\infty)}\right|^{2}}]\big){\rm {diag}}\{{\textbf{R}}^{a*}\}\!+\!{\rm {diag}}\{\Re[{{{\textbf{Q}}_{3}}(\infty)}]\}\nonumber \\
 & \!\simeq\!\mu(\sigma_{x}^{2}+\sigma_{q}^{2})\sigma_{v}^{2}{{\textbf{1}}_{2M}}+8k_{{\rm {TIQ}}}^{3}\sigma_{x}^{4}{{\textbf{p}}_{{\rm {IMD}}}^{\textmd{o}}}\label{diag156approx}
\end{align}
where the definition of ${{\textbf{p}}_{{\rm {IMD}}}^{\textmd{o}}}$ is given in \eqref{phIMD} in Appendix A. Upon replacing \eqref{diag156approx} into \eqref{evoofkinfty},
we arrive at
\begin{equation}
{{\bm{\kappa}}^{a}}(\infty)\!\!=\!\!\ \frac{\mu(\sigma_{v}^{2}+\sigma_{q}^{2}){{\textbf{1}}_{2M}}+8k_{{\rm {TIQ}}}^{3}\sigma_{x}^{4}{{\textbf{p}}_{{\rm {IMD}}}^{\textmd{o}}}}{{2(1-\mu\sigma_{x}^{2}-\mu M\sigma_{x}^{2})}}\label{kappaInftyLarge}
\end{equation}
According to the analysis in Appendix \ref{App:AppendixA}, we have
\begin{equation}
{\rm {Tr}}\big(\Re[{{{\textbf{Q}}_{3}}(\infty)}]\big)\!=\!4k_{{\rm {TIQ}}}^{3}\sigma_{x}^{6}[{{\left\Vert {{\textbf{h}}_{{\rm {IMD}}}^{\textmd{o}}}\right\Vert }^{2}}\!+\!{{\left\Vert {{\textbf{g}}_{{\rm {IMD}}}^{\textmd{o}}}\right\Vert }^{2}}]\label{traceE3E4}
\end{equation}
and from \eqref{un_vec},
\begin{equation}
E[{\left|{u(\infty)}\right|^{2}}]\!=\!\sigma_{v}^{2}\!+\!\sigma_{q}^{2}\!+\!6k_{{\rm {TIQ}}}^{3}\sigma_{x}^{6}[{{\left\Vert {{\textbf{h}}_{{\rm {IMD}}}^{\textmd{o}}}\right\Vert }^{2}}\!+\!{{\left\Vert {{\textbf{g}}_{{\rm {IMD}}}^{\textmd{o}}}\right\Vert }^{2}}]\label{uInfty}
\end{equation}
Therefore, upon replacing \eqref{kappaInftyLarge}-\eqref{uInfty} into \eqref{ACLMS_SS_MSE} and after a few algebraic manipulations, we arrive at
\begin{equation}
\begin{split} & J_{{\rm {high}}}^{a}(\infty)\!=\!\sigma_{v}^{2}+\sigma_{q}^{2}-2k_{{\rm {TIQ}}}^{3}\sigma_{x}^{6}[{{\left\Vert {{\textbf{h}}_{{\rm {IMD}}}^{\textmd{o}}}\right\Vert }^{2}}\!+\!{{\left\Vert {{\textbf{g}}_{{\rm {IMD}}}^{\textmd{o}}}\right\Vert }^{2}}]\\
 & +\frac{{\mu M(\sigma_{v}^{2}+\sigma_{q}^{2})\sigma_{x}^{2}+4k_{{\rm {TIQ}}}^{3}\sigma_{x}^{6}[{{\left\Vert {{\textbf{h}}_{{\rm {IMD}}}^{\textmd{o}}}\right\Vert }^{2}}\!+\!{{\left\Vert {{\textbf{g}}_{{\rm {IMD}}}^{\textmd{o}}}\right\Vert }^{2}}]}}{{1-\mu\sigma_{x}^{2}-\mu M\sigma_{x}^{2}}}
\end{split}
\label{MSEHighTx}
\end{equation}
to give the achievable SINR of augmented LMS based SI canceller in the case of high transmit power in \eqref{SINRAchvHighTxEqua} at the top of the next page.
\setcounter{TempEqCnt}{\value{equation}}
\begin{figure*}[t]
	\begin{equation}
	{\rm {SINR}}_{{\rm {high}}}\!=\!\frac{{p_{{\rm {x}}_{{\rm {SOI}}}}}}{J_{{\rm {high}}}^{a}(\infty)}\!=\!\frac{1\!-\!\mu(M\!+\!1)\sigma_{x}^{2}}{{\displaystyle {\left(\frac{1}{{\rm {SNR}}_{{\rm {req}}}}\!+\!\frac{\sigma_{q}^{2}}{{k_{{\rm {BB}}}}{k_{{\rm {LNA}}}}{k_{{\rm {TIQ}}}}{p_{{\rm {sen}}}}}\right)}\!+\!{\displaystyle \frac{2[1\!+\!\mu(M\!+\!1)\sigma_{x}^{2}]k_{{\rm {TIQ}}}^{2}\sigma_{x}^{6}({{\left\Vert {{\textbf{h}}_{{\rm {IMD}}}^{\textmd{o}}}\right\Vert }^{2}}\!+\!{{\left\Vert {{\textbf{g}}_{{\rm {IMD}}}^{\textmd{o}}}\right\Vert }^{2}})}{{k_{{\rm {BB}}}}{k_{{\rm {LNA}}}}{p_{{\rm {sen}}}}}}}}\label{SINRAchvHighTxEqua}
	\end{equation}
	\hrulefill{}
\end{figure*}
\setcounter{equation}{37}

\textit{Remark 3:} From \eqref{SINRAchv} and \eqref{SINRAchvHighTxEqua}, we observe that, for both low and high transmit powers, the achievable SINRs of  augmented LMS at the steady state are monotonically decreasing functions of the step-size $\mu$, the end-to-end channel impulse response length $M$, and the {SI power} $\sigma_{x}^{2}$. Particularly, in the case of high transmit powers, the achievable SINR of  augmented LMS is impaired by the non-negligible IMD SI $\textbf{x}_{{\rm {IMD}}}(n)$ and its image $\textbf{x}_{{\rm {IMD}}}^{*}(n)$, as well as their associated end-to-end channel impulse responses, and its degradation in SINR becomes more severe with an increase in transmit power.
\section{Proposed Augmented Nonlinear LMS based SI Canceller and Its Performance Analysis}
\label{ANCLMS} From the mean and mean square analysis in Section
\ref{pfmAna}, it is clear that, for high transmit powers,
the third order IMD component $\textbf{x}_{{\rm {IMD}}}(n)$, produced by PA distortion, may exceed
the thermal noise floor, or even become stronger than the signal of
interest $x_{{\rm {SOI}}}(n)$. This leads to bias as well as suboptimality in
SINR performance of the conventional  augmented LMS based SI canceller. To
address these issues, it is desirable to design an adaptive SI canceller which suppresses the SI, image SI
and IMD components simultaneously. Upon revisiting the vectorized signal model
in \eqref{dn_vec}  and \eqref{dn_aug}, instead of
treating the IMD components ${\textbf{x}}_{{\rm {IMD}}}(n)$ and ${\textbf{x}}_{{\rm {IMD}}}^{*}(n)$
as a part of the aggregated interference $u(n)$, we can concatenate them with
the SI component ${\textbf{x}}(n)$ and its image counterpart ${\textbf{x}}^{*}(n)$
to form a $(2M+2N)\times1$ vector ${\textbf{x}}^{b}(n)$, given by
\begin{equation}\label{vector_x_b}
{{\textbf{x}}^{b}}(n)={[{{\textbf{x}}^{T}}(n),{\textbf{x}}_{{\rm {IMD}}}^{T}(n),{{\textbf{x}}^{H}}(n),{\textbf{x}}_{{\rm {IMD}}}^{H}(n)]^{T}}
\end{equation}
Note that, compared with the augmented SI vector ${{\textbf{x}}^{a}}(n)$ in \eqref{dn_aug}, vector ${\textbf{x}}^{b}(n)$ is no longer simply widely linear in ${\textbf{x}}(n)$. In fact, by defining
\begin{equation*}
{{\textbf{x}}^{c}}(n)=[{{\textbf{x}}^{T}}(n),{\textbf{x}}^{dT}(n)]^T\label{eq:xcn}
\end{equation*}
where ${\textbf{x}}^{d}(n)={[x(n),x(n-1),\ldots,x(n-N+1)]^{T}}$ contains the first $N$ elements in ${\textbf{x}}(n)$, ${{\textbf{x}}^{b}}(n)$ in \eqref{vector_x_b} can be represented in a widely nonlinear form in ${{\textbf{x}}^{c}}(n)$ (or equivalently, in ${\textbf{x}}(n)$) as \cite{Xia2011,Xu2015}
%
\begin{align*}
{\textbf{x}}^{b}(n) = {\bm{\Psi}}^{T}\big([{\textbf{x}}^{cT}(n), {\textbf{x}}^{cH}(n)]\big)
\end{align*}
where ${\bm{\Psi}}(\cdot)$ is a vectorized nonlinear function, Based on \eqref{xIMDn}, ${\bm{\Psi}}\big({\textbf{x}}^{c}(n)\big)$ can be expressed element by element as
\begin{align*}\label{eq:Psi}
&{\bm{\Psi}}\big({\textbf{x}}^{c}(n)\big)\!\!=\!\![{\psi}_{1}\big({x(n)}\big),{\psi}_{2}\big({x(n\!-\!1)}\big),\dots,{\psi}_{M}\big({x(n\!-\!M\!+\!1)}\big),\nonumber\\
&{\psi}_{M\!+\!1}\big({x(n)}\big),{\psi}_{M\!+\!2}\big({x(n\!-\!1)}\big),\dots,{\psi}_{M\!+\!N}\big({x(n\!-\!N\!+\!1)}\big)]^T
\end{align*}
where
\begin{equation*}\label{eq:psix}
\!{\psi}_i(x) \!=\! \left\{ {\begin{array}{*{20}{c}}
	{\hspace{-3mm}x,}&{\hspace{-20mm}\!\!i=1,2,\dots,M} \\
	{k_{{\rm {TIQ}}}^{3/2}\left| x \right|^2x,}&\!\!{i=M\!+\!1,M\!+\!2,\dots,M\!+\!N}
	\end{array}} \right.
\end{equation*}
Based on the above discussion, we refer to ${{\textbf{x}}^{b}}(n)$ as the augmented nonlinear SI vector, and for SI cancellation of FD transceivers in the presence of PA nonlinearity, we should take into account $\textbf{x}^{b}(n)$ to give a model-fit widely nonlinear relation between the observed signal $d(n)$ and its corresponding SI waveform $x(n)$ in the form
\begin{equation}
d(n)={\textbf{x}}^{bT}(n){{\textbf{w}}^{b\textmd{o}}}+v(n)+q(n)
\end{equation}
where ${{\textbf{w}}^{b\textmd{o}}}={[{\textbf{h}}^{{\textmd{o}}T}, {\textbf{h}}_{{\rm {IMD}}}^{{\textmd{o}}T}, {\textbf{g}}^{{\textmd{o}}T}, {\textbf{g}}_{{\rm {IMD}}}^{{\textmd{o}}T}]^{T}}$
is the $(2M+2N)\times1$ sufficient-length end-to-end filter impulse response of a FD transceiver. Similar to the conventional augmented LMS, the proposed augmented nonlinear LMS based SI canceller aims to estimate ${{\textbf{w}}^{b\textmd{o}}}$ by minimizing a mean square error cost function $J^{b}(n)$, defined as
\begin{equation}
J^{b}(n)=E[{\left|{{e^{b}}(n)}\right|^{2}}]=E[{{e^{b}}(n)}{{e^{b*}}(n)}]\label{MSEDef2}
\end{equation}
where the instantaneous residual SI $e^{b}(n)$ is given by
\begin{equation}
e^{b}(n)=d(n)-{\textbf{x}}^{bT}(n){{\textbf{w}}^{b}}(n)\label{ebn}
\end{equation}
and governs the update of the weight vector ${{\textbf{w}}^{b}}(n)$
as
\begin{equation}
{{\textbf{w}}^{b}}(n\!+\!1)={{\textbf{w}}^{b}}(n)+\mu e^{b}(n){\textbf{x}}^{b*}(n)\label{wnIterJ}
\end{equation}
\subsection{Mean Convergence Analysis}
Upon introducing the $(2M+2N)\times1$ weight error vector
\begin{eqnarray}
{\widetilde{\textbf{w}}}^{b}(n)={\textbf{w}}^{b}(n)-{{\textbf{w}}^{b\textmd{o}}}\label{eqwb}
\end{eqnarray}
the output residual SI ${{e^{b}}(n)}$ in \eqref{ebn} becomes
\begin{equation}
e^{b}(n)=v(n)+q(n)-{\textbf{x}}^{bT}(n){\widetilde{\textbf{w}}}^{b}(n)\label{en2J}
\end{equation}
and the recursion of ${\widetilde{\textbf{w}}}^{b}(n)$ now obeys
\begin{align}
{\widetilde{\textbf{w}}}^{b}(n\!+\!1) & =[{\textbf{I}}_{2M+2N}-\mu{\textbf{x}}^{b*}(n){\textbf{x}}^{bT}(n)]{\widetilde{\textbf{w}}}^{b}(n)\nonumber \\
 & +\mu[v(n)+q(n)]{\textbf{x}}^{b*}(n)\label{misalgnIterb}
\end{align}
Upon applying the expectation operator $E[\cdot]$ on both sides of \eqref{misalgnIterb} and using the standard independence assumptions introduced in Section \ref{meanConvAna}, we arrive at
\begin{align}
E[{\widetilde{\textbf{w}}}^{b}(n\!+\!1)] & \!=\!({\textbf{I}}_{2M+2N}\!-\!\mu{\textbf{R}}^{b*})E[{\widetilde{\textbf{w}}}^{b}(n)]\label{misalgnIterExpJ}
\end{align}
where ${{\textbf{R}}^{b}}=E[{\textbf{x}}^{b}(n){\textbf{x}}^{bH}(n)]$ is the covariance matrix of the augmented nonlinear SI input vector ${\textbf{x}}^{b}(n)$. Based on \eqref{misalgnIterExpJ}, the step-size $\mu$ which guarantees the convergence of the proposed  augmented nonlinear LMS in
the mean sense should satisfy
\begin{equation*}
\left|{1-\mu{\lambda_{k}^{b}}}\right|<1,\quad\quad k=1,\ldots,2M+2N
\end{equation*}
where $\lambda_{k}^{b}$ are the eigenvalues of ${{\textbf{R}}^{b}}$.
Note that although ${{\textbf{R}}^{b}}$ is Hermitian, its positive-definiteness
is not always guaranteed due to the non-Gaussianity of ${{\textbf{x}}^{b}}(n)$.
To investigate this issue, we shall further decompose ${{\textbf{R}}^{b}}$
as
\begin{equation*}
{{\textbf{R}}^{b}}\!=\!E[{\textbf{x}}^{b}(n){\textbf{x}}^{bH}(n)]\!=\!{\left[{\begin{array}{cc}
{{\textbf{R}}_{0}^{b}} & {\textbf{0}}\\
{\textbf{0}} & {{\textbf{R}}_{0}^{b}}
\end{array}}\right]}\label{RxJ}
\end{equation*}
where
\begin{align*}
{{\textbf{R}}_{0}^{b}}\!\!&=\!\!E[{{\textbf{x}}^{e}}(n){{\textbf{x}}^{eH}}(n)]\!\!=\!\!{\left[{\begin{array}{cc}{E[{\left|x(n)\right|^{2}}]{{\textbf{I}}_{M}}} \!\!&\!\! {{\bm {\Omega}}^{T}}\\{\bm {\Omega}} \!\!&\!\! {{k_{{\rm{TIQ}}}^{3}}E[{\left|x(n)\right|^{6}}]{{\textbf{I}}_{N}}}\end{array}}\right]}
\end{align*}

\[
{\bm {\Omega}}={[\begin{array}{cc}
{{\textbf{R}}^{d}} & {{\textbf{0}}_{N\times(M-N)}}\end{array}]}
\]
\[
{{\textbf{R}}^{d}}\!=\!E[{\textbf{x}}^{d}(n){\textbf{x}}_{{\rm {IMD}}}^{H}(n)]\!=\!{{k_{{\rm {TIQ}}}^{3/2}}E[{\left|x(n)\right|^{4}}]{{\textbf{I}}_{N}}}
\]
in which ${{\textbf{x}}^{e}}(n)=[{{\textbf{x}}^{T}}(n),{\textbf{x}}_{{\rm {IMD}}}^{T}(n)]^T$. Now, by solving $\det[{{\textbf{R}}^{b}}-\lambda_{k}^{b}{{\textbf{I}}_{2M+2N}}]=0$ and after some algebraic manipulations, we arrive at
\begin{align}
{\lambda_{1}^{b}}\!\! & =  \!\!E[{\left|x(n)\right|^{2}}]=\sigma_{x}^{2}\nonumber \\
{\lambda_{2,3}^{b}}\!\! & =  \!\!\frac{1}{2}\Big\{ E[{\left|x(n)\right|^{2}}]+k_{{\rm {TIQ}}}^{3}E[{\left|x(n)\right|^{6}}]\nonumber \\
\!\! & \pm  \!\!\sqrt{\big(E[{{\left|x(n)\right|}^{2}}]\!+\!k_{{\rm{TIQ}}}^{3}E[{{\left|x(n)\right|}^{6}}]\big)^{2}\!-\!k_{{\rm {TIQ}}}^{3}\big({E}[{{\left|x(n)\right|}^{4}}]\big)^{2}}\Big\}\nonumber \\
\!\! & =  \!\!\frac{{\sigma_{x}^{2}\!+\!6{k_{{\rm {TIQ}}}^{3}}\sigma_{x}^{6}\!\pm\!\sigma_{x}^2\sqrt{{1\!-\!2{k_{{\rm {TIQ}}}^{3}}\sigma_{x}^{4}}\!+\!36{k_{{\rm {TIQ}}}^{6}}\sigma_{x}^{8}}}}{2}\label{eigenvalue}
\end{align}
%
where the algebraic multiplicities of ${\lambda_{1}^{b}}$, ${\lambda_{2}^{b}}$
and ${\lambda_{3}^{b}}$ are respectively $2M-2N$, $2N$ and $2N$.

\textit{Remark 4:} The covariance matrix ${{\textbf{R}}^{b}}$ is
positive-definite since all the eigenvalues are positive. It is also
worth noting that, although $x(n)$ is considered to be i.i.d. Gaussian, the positive definiteness of ${{\textbf{R}}^{b}}$ is
still valid as long as $x(n)$ comes from any i.i.d. second order
circular constellation, e.g., QPSK or M-QAM. Note that in \eqref{eigenvalue}, the largest eigenvalue of ${{\textbf{R}}^{b}}$
is ${\lambda_{2}^{b}}$, and therefore, the mean convergence bound on
the step-size $\mu$ is given by
\begin{align}
 & 0<\mu<\frac{2}{{{\lambda_{\max}}[{{\textbf{R}}^{b}}]}}\nonumber \\
 & =\frac{4}{{\sigma_{x}^{2}\!+\!6{k_{{\rm {TIQ}}}^{3}}\sigma_{x}^{6}\!+\!\sigma_{x}^2\sqrt{{1\!-\!2{k_{{\rm {TIQ}}}^{3}}\sigma_{x}^{4}}\!+\!36{k_{{\rm {TIQ}}}^{6}}\sigma_{x}^{8}}}}\label{stepSizeMeanCvgJ}
\end{align}
which enables the proposed  augmented nonlinear LMS based SI canceller to asymptotically
achieve unbiased estimation of the optimal weight vector ${{\textbf{w}}^{b\textmd{o}}}$, indicated by $E[{\widetilde{\textbf{w}}}^{b}(\infty)]=\textbf{0}$ from \eqref{misalgnIterExpJ}, and independent on whether the transmit power of an FD transceiver is low or high.
\subsection{Mean Square Convergence Analysis}
Upon taking \eqref{en2J} into \eqref{MSEDef2}, and again employing the standard independence assumptions, the MSE of the proposed  augmented nonlinear LMS based SI canceller can be further evaluated as
\begin{equation}
{J^{b}}(n)\!=\!{\rm {Tr}}[{{\textbf{R}}^{b}}{{\textbf{K}}^{b}}(n)]\!+\!\sigma_{v}^{2}+\sigma_{q}^{2}\label{MSEb}
\end{equation}
where ${{\textbf{K}}^{b}}(n)=E[{{\widetilde{\textbf{w}}}^{b}}(n){\widetilde{\textbf{w}}}^{bH}(n)]$
is the covariance matrix of the weight error vector ${{\widetilde{\textbf{w}}}^{b}}(n)$.
To analyze its evolution, we first multiply both sides of \eqref{misalgnIterb}
by ${\widetilde{\textbf{w}}}^{bH}(n)$, and apply the statistical
expectation operator $E[\cdot]$, to give
\begin{align}
&{{\textbf{K}}^{b}}(n\!+\!1)={{\textbf{K}}^{b}}(n)+{\mu^{2}}(\sigma_{v}^{2}+\sigma_{q}^{2}){\textbf{R}}^{b}\nonumber\\
&{-\mu E[{\textbf{x}}^{b*}(n){\textbf{x}}^{bT}(n){{\widetilde{\textbf{w}}}^{b}}(n){{\widetilde{\textbf{w}}}^{bH}}(n)]}\nonumber\\
&{-\mu E[{{\widetilde{\textbf{w}}}^{b}}(n){{\widetilde{\textbf{w}}}^{bH}}(n){\textbf{x}}^{b*}(n){\textbf{x}}^{bT}(n)]}\nonumber\\
&{+{\mu^{2}}E[{\textbf{x}}^{b*}(n){\textbf{x}}^{bT}(n){{\widetilde{\textbf{w}}}^{b}}(n){{\widetilde{\textbf{w}}}^{bH}}(n){\textbf{x}}^{b*}(n){\textbf{x}}^{bT}(n)]}\label{misalgn2ndOrderJ}
\end{align}
Since the augmented nonlinear SI vector ${\textbf{x}}^{b}(n)$ in \eqref{vector_x_b} is non-Gaussian, the Gaussian fourth
order moment factorizing theorem used in Section \ref{pfmAna} is
no longer applicable to evaluate the last term on the RHS of \eqref{misalgn2ndOrderJ}, and we therefore resort to matrix vectorization instead of matrix diagonalization to evaluate the mean square convergence behavior of the proposed  augmented nonlinear LMS
\cite{AlNaffouri2003,sayed2011adaptive}. By using the following matrix
vectorization lemma for arbitrary matrices $\{\textbf{A},\textbf{B},\textbf{C}\}$
\[
{\rm {vec}}\{\textbf{ABC}\}=({{\textbf{C}}^{T}}\otimes{\textbf{A}}){\rm {vec}}\{\textbf{B}\}
\]
it is straightforward to verify that the recursion for ${{\textbf{K}}^{b}}(n)$ in \eqref{misalgn2ndOrderJ} can be transformed into a linear vector relation as
\begin{align}
{\rm {vec}}\{{\textbf{K}}^{b}(n\!+\!1)\} & =  (\underbrace{{{\textbf{I}}_{2M+2N}}-\mu{\textbf{S}}+{\mu^{2}}{\textbf{T}}}_{{{\textbf{F}}^{b}}}){\rm {vec}}\{{{\textbf{K}}^{b}}(n)\}\nonumber \\
 & +  {\mu^{2}}(\sigma_{v}^{2}+\sigma_{q}^{2}){\rm {vec}}\{{\textbf{R}}^{b}\}\label{vecKbIter}
\end{align}
where
\begin{eqnarray*}
{\textbf{S}} & = & {{\textbf{I}}_{2M+2N}}\otimes{\textbf{R}}^{b}+{\textbf{R}}^{b}\otimes{{\textbf{I}}_{2M+2N}}\\
{\textbf{T}} & = & E\big[\big({{\textbf{x}}^{b}}(n){\textbf{x}}^{bH}(n)\big)\otimes\big({\textbf{x}}^{b*}(n){\textbf{x}}^{bT}(n)\big)\big]
\end{eqnarray*}
The condition on the step-size $\mu$ to guarantee the convergence
of ${{\textbf{K}}^{b}}(n)$ now relies on $\left|{\lambda[{{\textbf{F}}^{b}}]}\right|<1$.
It has been proven in \cite{sayed2011adaptive} that for positive
definite ${\textbf{R}}^{b}$ and ${\textbf{S}}$ and nonnegative definite ${\textbf{T}}$, the stability of the recursion in \eqref{vecKbIter}
is guaranteed when
\begin{equation}
0\!<\!\mu\!<\!\min\left\{ {\frac{1}{{{\lambda_{\max}}[{{\textbf{S}}^{-1}}{\textbf{T}}]}}, \frac{1}{\lambda_{\max}[{\bm{\Gamma}}]}}\right\} \label{Mean_Square_ANCLMS}
\end{equation}
where
\begin{equation*}
{\bm{\Gamma}}={\left[{\begin{array}{cc}
{\frac{{\textbf{S}}}{2}} & {-\frac{{\textbf{T}}}{2}}\\
{{\textbf{I}}_{{{(2M+2N)}^{2}}}} & {\textbf{0}}
\end{array}}\right]}
\end{equation*}
\subsection{Steady State Analysis}
Suppose the step-size $\mu$ is chosen to ensure the mean square stability
of the proposed  augmented nonlinear LMS. Then, from \eqref{MSEb}, its steady-state MSE ${J^{b}}(\infty)$
can be expressed as
\begin{equation}
{J^{b}}(\infty)\!=\!{\rm {Tr}}[{{\textbf{R}}^{b}}{{\textbf{K}}^{b}}(\infty)]\!+\!\sigma_{v}^{2}+\sigma_{q}^{2}\label{MSEJ}
\end{equation}
where ${{\textbf{K}}^{b}}(\infty)$ can be evaluated from \eqref{vecKbIter}
as
\begin{eqnarray}
{{\textbf{K}}^{b}}(\infty)\!\!\!\! & = & \!\!\!\!{{\rm {vec}}^{-1}}\{{\mu^{2}}(\sigma_{v}^{2}\!+\!\sigma_{q}^{2}){({{\textbf{I}}_{2M+2N}}\!-\!{{\textbf{F}}^{b}})^{-1}}{\rm {vec}}\{{\textbf{R}}^{b}\}\}\nonumber \\
\!\!\!\!\!\! & = & \!\!\!\!\!\!{{\rm {vec}}^{-1}}\{{\mu^{2}}(\sigma_{v}^{2}\!+\!\sigma_{q}^{2}){(\mu\textbf{S}\!-\!\mu^{2}\textbf{T})^{-1}}{\rm {vec}}\{{\textbf{R}}^{b}\}\}\label{vecKb}
\end{eqnarray}
Upon substituting \eqref{vecKb} into \eqref{MSEJ}, we have
\begin{eqnarray*}
\!\!\!\!\!\!\!\!{J^{b}}(\infty)\!\!\!\!&=&\!\!\!\!(\sigma_{v}^{2}\!+\!\sigma_{q}^{2})\nonumber\\
\!\!\!\!&\cdot&\!\!\!\!\Big(1\!\!+\!\!{\mu^{2}}{\rm {Tr}}\!\big[{\textbf{R}}^{b}{\rm {vec}}{^{-1}}\!\{{(\mu\textbf{S}\!-\!\mu^{2}\textbf{T})^{-1}}{\rm {vec}}\!\{{\textbf{R}}^{b}\}\}\big]\Big)\label{MSEANCLMS}
\end{eqnarray*}
Due to the existence of the fourth order moment matrix ${\textbf{T}}$,
a detailed evaluation of the steady state MSE ${J^{b}}(\infty)$ is much more difficult than
that of the standard augmented LMS. However, as shown in \eqref{vecKbIter}, since matrix ${\textbf{T}}$ is multiplied by ${\mu}^{2}$ within the matrix ${{\textbf{F}}^{b}}$,
then for a sufficient small step-size $\mu$, we  can neglect the term
$\mu^{2}{\textbf{T}}$ in ${{\textbf{F}}^{b}}$ \cite{AlNaffouri2003,sayed2011adaptive}. In this way, the standard eigenvalue decomposition (EVD) of ${{\textbf{R}}^{b}}$
gives ${{\textbf{R}}^{b}}={\bf {U}}{{\bm {\Lambda}}^{b}}{\bf {U}}^{H}$,
where ${\bf {U}}$ is a unitary matrix and ${{\bm {\Lambda}}^{b}}={\rm {diag}}\{\lambda_{1}^{b}, \lambda_{2}^{b}, \ldots, \lambda_{2M+2N}^{b}\}$
is a diagonal matrix comprising of the eigenvalues of ${{\textbf{R}}^{b}}$,
from \eqref{misalgn2ndOrderJ}, to give
\begin{equation*}
{{\widetilde{\textbf{K}}}^{b}}(n\!+\!1)\!=\!{\widetilde{\textbf{K}}^{b}}(n)\!+\!{\mu^{2}}(\sigma_{v}^{2}\!+\!\sigma_{q}^{2}){{\bm{\Lambda}}^{b}}\!-\!\mu{{\bm{\Lambda}}^{b}}{\widetilde{\textbf{K}}^{b}}(n)\!-\!\mu{\widetilde{\textbf{K}}^{b}}(n){{\bm{\Lambda}}^{b}}\label{misalgn2ndOrderJreduced}
\end{equation*}
where ${{\widetilde{\textbf{K}}}^{b}}(n)={{\bf {U}}^{H}}{{\textbf{K}}^{b}}(n){\bf {U}}$,
and its steady-state value ${{\widetilde{\textbf{K}}}^{b}}(\infty)$
can be obtained as
\begin{equation*}
{{\widetilde{\textbf{K}}}^{b}}(\infty)=\frac{\mu(\sigma_{v}^{2}+\sigma_{q}^{2})}{2}{{\textbf{I}}_{2M+2N}}\label{ss_tilde_k}
\end{equation*}
After substituting into \eqref{MSEJ}, we obtain an approximated steady-state
MSE, denoted by ${J_{{\rm {ap}}}^{b}(\infty)}$, in the form
\begin{eqnarray}
{J_{{\rm {ap}}}^{b}(\infty)}\!\!\!\!  & = & \!\!\!\! {\rm {Tr}}[{\bf {U}}{{\bm{\Lambda}}^{b}}{{\widetilde{\textbf{K}}}^{b}}(\infty){{\bf {U}}^{H}}]+\sigma_{v}^{2}+\sigma_{q}^{2}\nonumber \\
\!\!\!\!  & = &  \!\!\!\!(\sigma_{v}^{2}\!+\!\sigma_{q}^{2})[\mu(M\sigma_{x}^{2}\!+\!6N{k_{{\rm {TIQ}}}^{3}}\sigma_{x}^{6})\!+\!1]\label{MSEANCLMSAP}
\end{eqnarray}
Now from \eqref{sigmav}, \eqref{pxSOI} and \eqref{MSEANCLMSAP}, the achievable SINR of the proposed  augmented nonlinear LMS can be evaluated as in \eqref{SINRAchvANCLMSEqua}
at the top of the next page.
\setcounter{TempEqCnt}{\value{equation}} 
\begin{figure*}[t]
	\begin{equation}
	{\rm {SINR}}_{{\rm {Proposed}}}\!=\!\frac{{p_{{\rm {x}}_{{\rm {SOI}}}}}}{{J_{{\rm {ap}}}^{b}(\infty)}}\!=\!\frac{1}{{\Big({\displaystyle \frac{1}{{{\rm {SNR}}_{{\rm {req}}}}}\!+\!{\displaystyle \frac{{\sigma_{q}^{2}}}{{{k_{{\rm {BB}}}}{k_{{\rm {LNA}}}}{k_{{\rm {TIQ}}}}{p_{{\rm {sen}}}}}}\Big)[1\!+\!\mu(M\sigma_{x}^{2}\!+\!6Nk_{{\rm {TIQ}}}^{3}\sigma_{x}^{6})]}}}}\label{SINRAchvANCLMSEqua}
	\end{equation}
	\hrulefill{}
\end{figure*}
\setcounter{equation}{56}

\textit{Remark 5:} The achievable SINR of the proposed  augmented nonlinear LMS based SI canceller in \eqref{SINRAchvANCLMSEqua} is a monotonically decreasing function of the step-size $\mu$, the length of SI channel impulse response $M$ and that of IMD SI channel impulse response $N$, the transmitter mixer gain $k_{{\rm {TIQ}}}$, and the SI power $\sigma_{x}^{2}$. Moreover, in the situations of high transmit powers, owing to the model fitting advantage, the optimality of the proposed  augmented nonlinear LMS for nonlinear SI cancellation can be also observed, since its achievable SINR is no longer impaired by the IMD channel impulse responses, that is, $\left\Vert{{\textbf{h}}_{{\rm {IMD}}}^{{\textmd{o}}}}\right\Vert^{2}$
and $\left\Vert{{\textbf{g}}_{{\rm {IMD}}}^{{\textmd{o}}}}\right\Vert^{2}$, which, however, remain the by-products of the undermodeling problem encountered by the conventional  augmented LMS based SI canceller for FD transceivers in the joint presence of PA nonlinearity and frequency-dependent IQ imbalance.
\begin{figure*}[t!]
\vspace{0.2cm}
\centering \subfigure[ Augmented LMS]{ \label{meanbehaviorfig:ACLMS} 
\begin{minipage}[b]{0.48\textwidth}%
 \centering \includegraphics[width=1\textwidth]{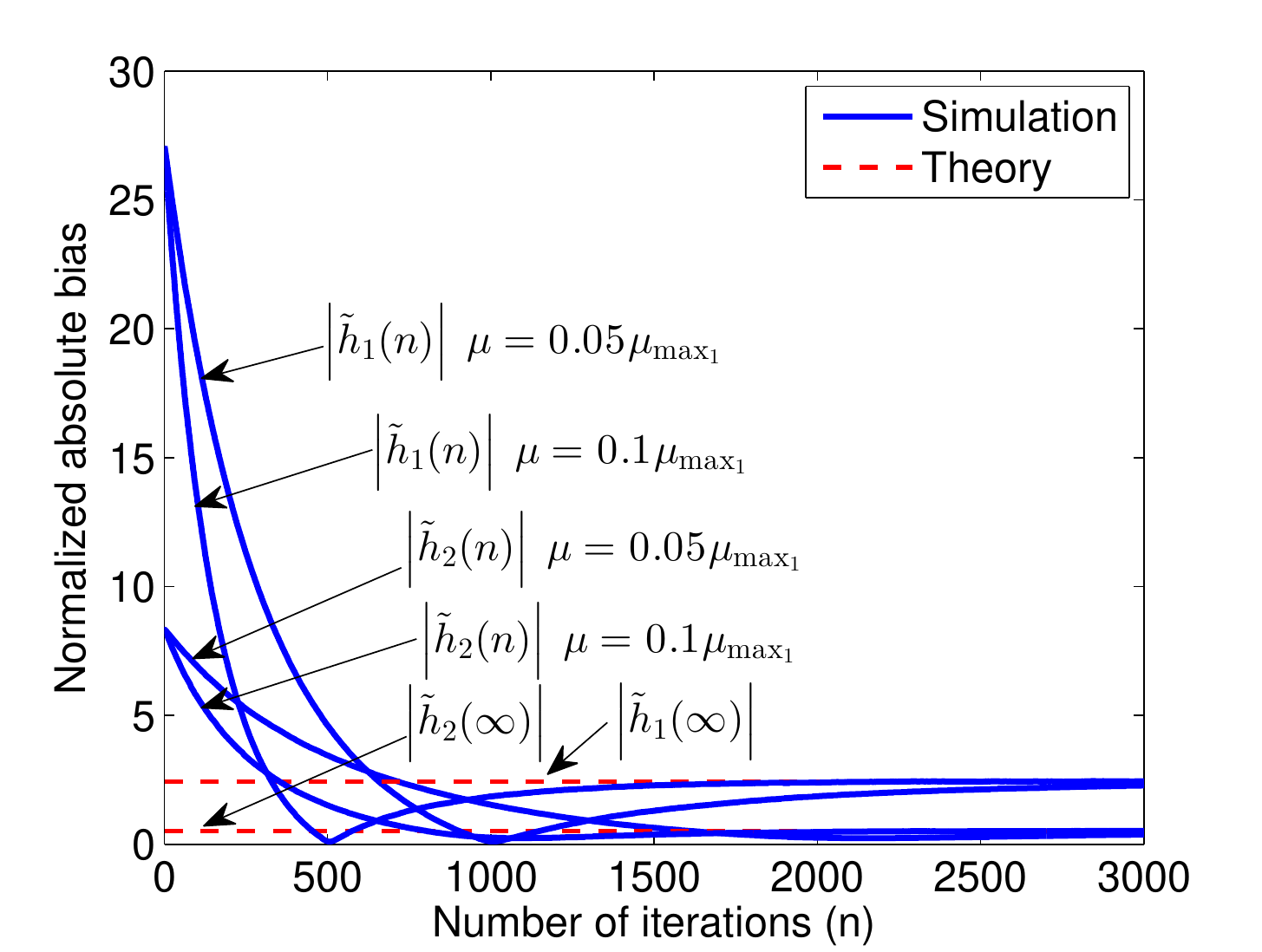} %

\end{minipage}}\subfigure[ Augmented nonlinear LMS]{ \label{meanbehaviorfig:ANCLMS} 
\begin{minipage}[b]{0.48\textwidth}%
 \centering \includegraphics[width=1\textwidth]{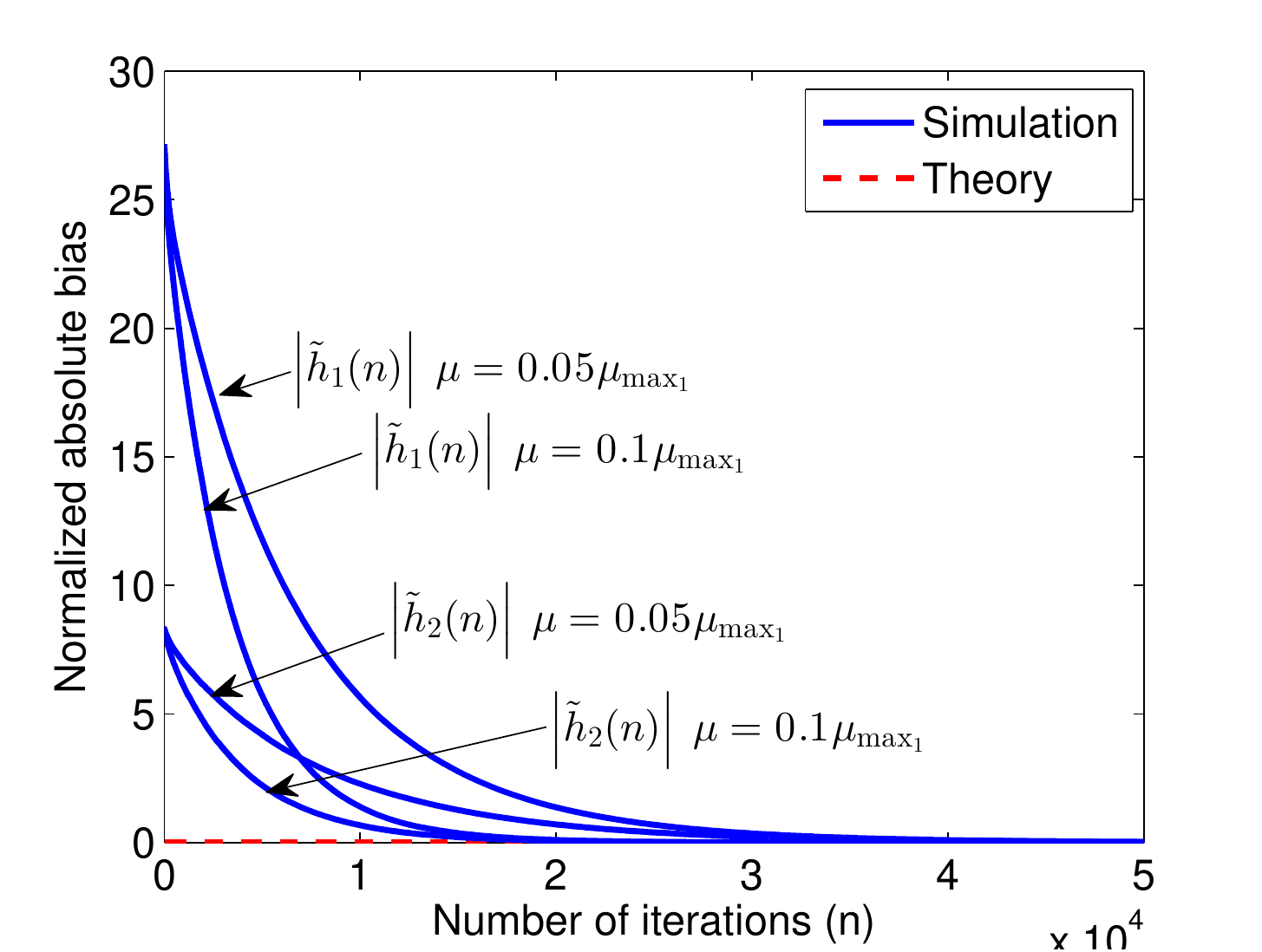} %
\end{minipage}} \caption{Normalized absolute values of two representative error coefficients
${{\widetilde{h}}_{1}}(n)$ and ${{\widetilde{h}}_{2}}(n)$ for a
Type 2 FD transceiver by using (a)  augmented LMS and (b)  augmented nonlinear LMS, where $\mu\in\{0.05\mu_{{\rm {max_1}}},0.1\mu_{{\rm {max_1}}}\}$.}
\label{meanbehaviorfig}
\vspace{0.2cm}
\end{figure*}

%
\section{Computer Simulations}\label{sim}
%
\vspace{0.2cm}
\begin{figure*}[t!]
	\centering \subfigure[Achievable SINR]{ \label{DigAtt_SINR:SINR}
		%
		\begin{minipage}[b]{0.48\textwidth}%
			\centering \includegraphics[width=1\textwidth]{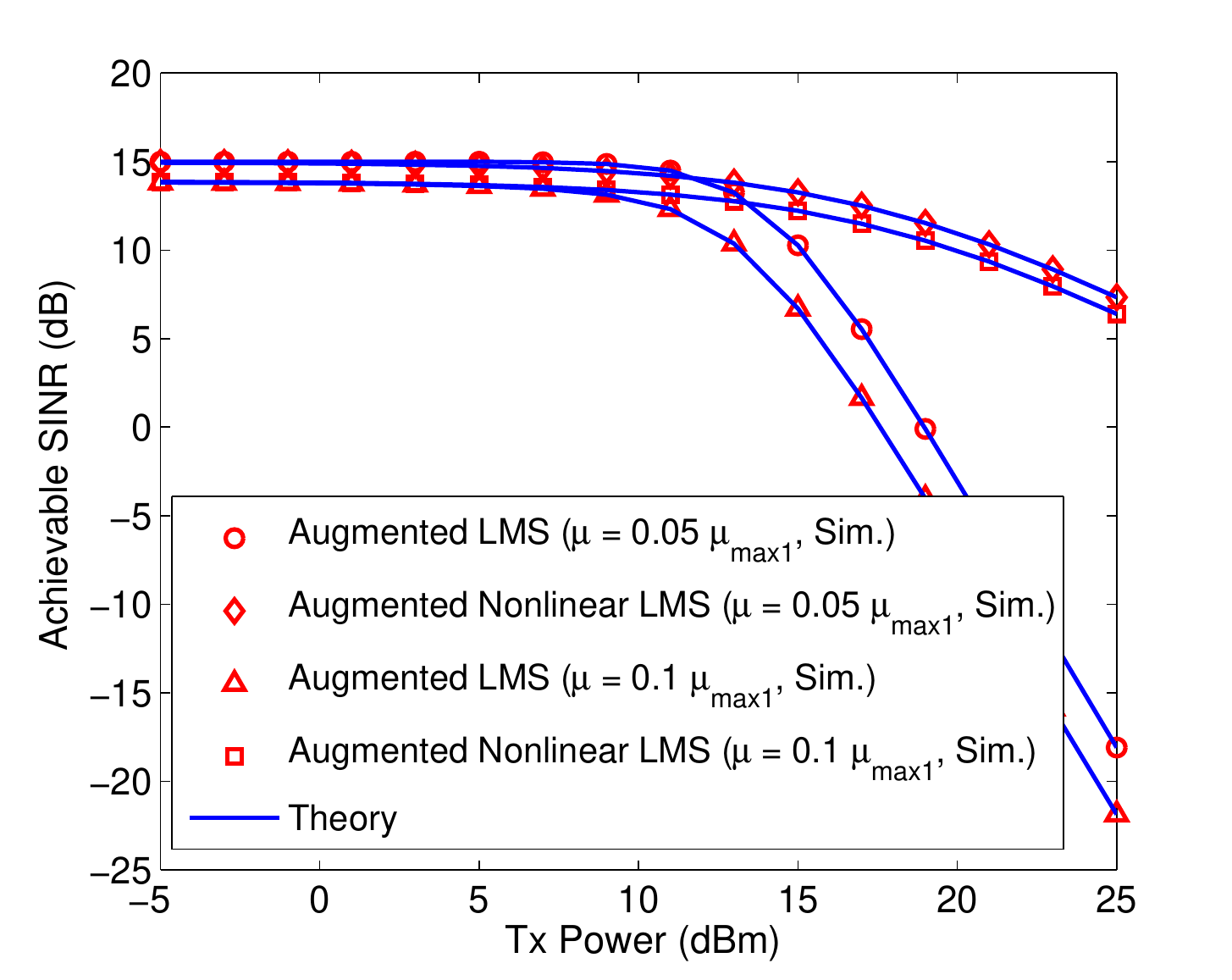} %
	\end{minipage}}\hfill{}\subfigure[Digital attenuation]{ \label{DigAtt_SINR:DigAtt}
		%
		\begin{minipage}[b]{0.48\textwidth}%
			\centering \includegraphics[width=1\textwidth]{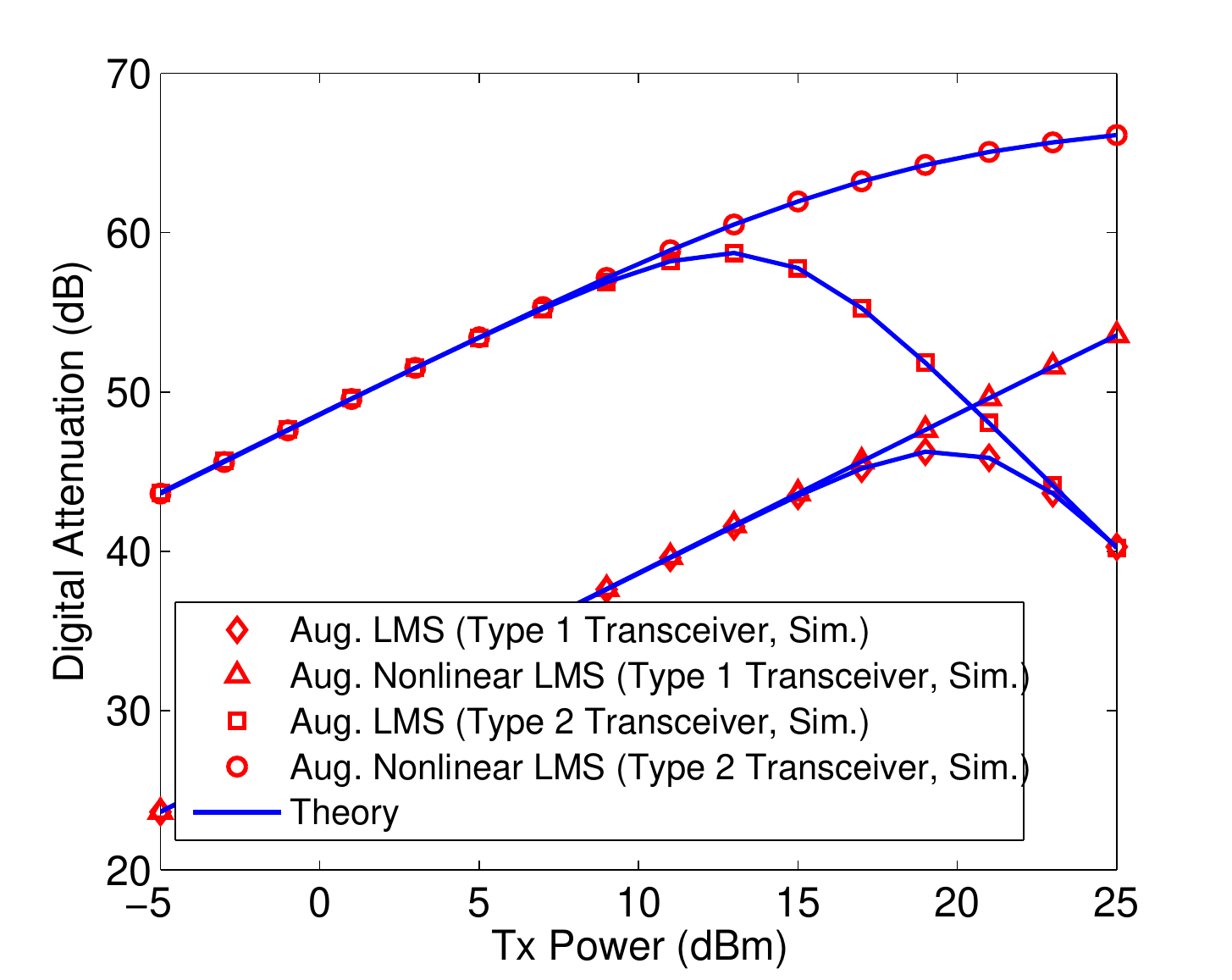} %
	\end{minipage}} \caption{Comparison of the theoretical and simulated steady-state mean square
		performances of both the conventional  augmented LMS and the proposed  augmented nonlinear LMS,
		measured in terms of (a) Achievable SINR, and (b) Digital attenuation.}
	\label{DigAtt_SINR} 
\end{figure*}
\vspace{0.2cm}
\begin{figure*}[t!]
	\centering \subfigure[Condition number of ${{\textbf{R}}^{b}}$ against $\sigma_{x}^2$ and $k_{\rm TIQ}$]{ \label{ANCLMSvspANCLMS:CondNo} 
		\begin{minipage}[b]{0.48\textwidth}%
			\centering \includegraphics[width=1\textwidth]{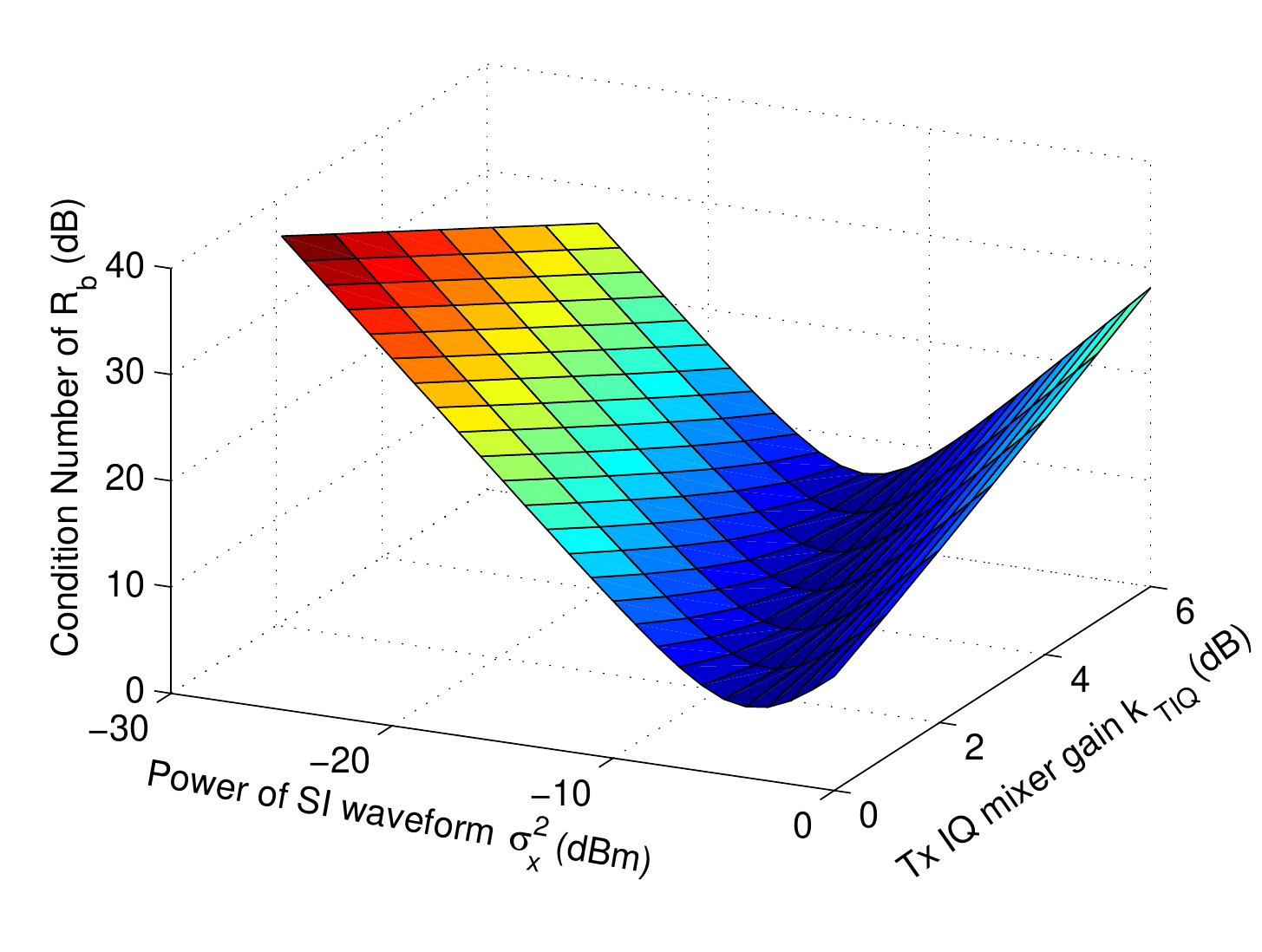} %
	\end{minipage}}\hfill{}\subfigure[SINR evolution of all the considered SI cancellers]{ \label{ANCLMSvspANCLMS:cvg-state} 
		\begin{minipage}[b]{0.48\textwidth}%
			\centering \includegraphics[width=1\textwidth]{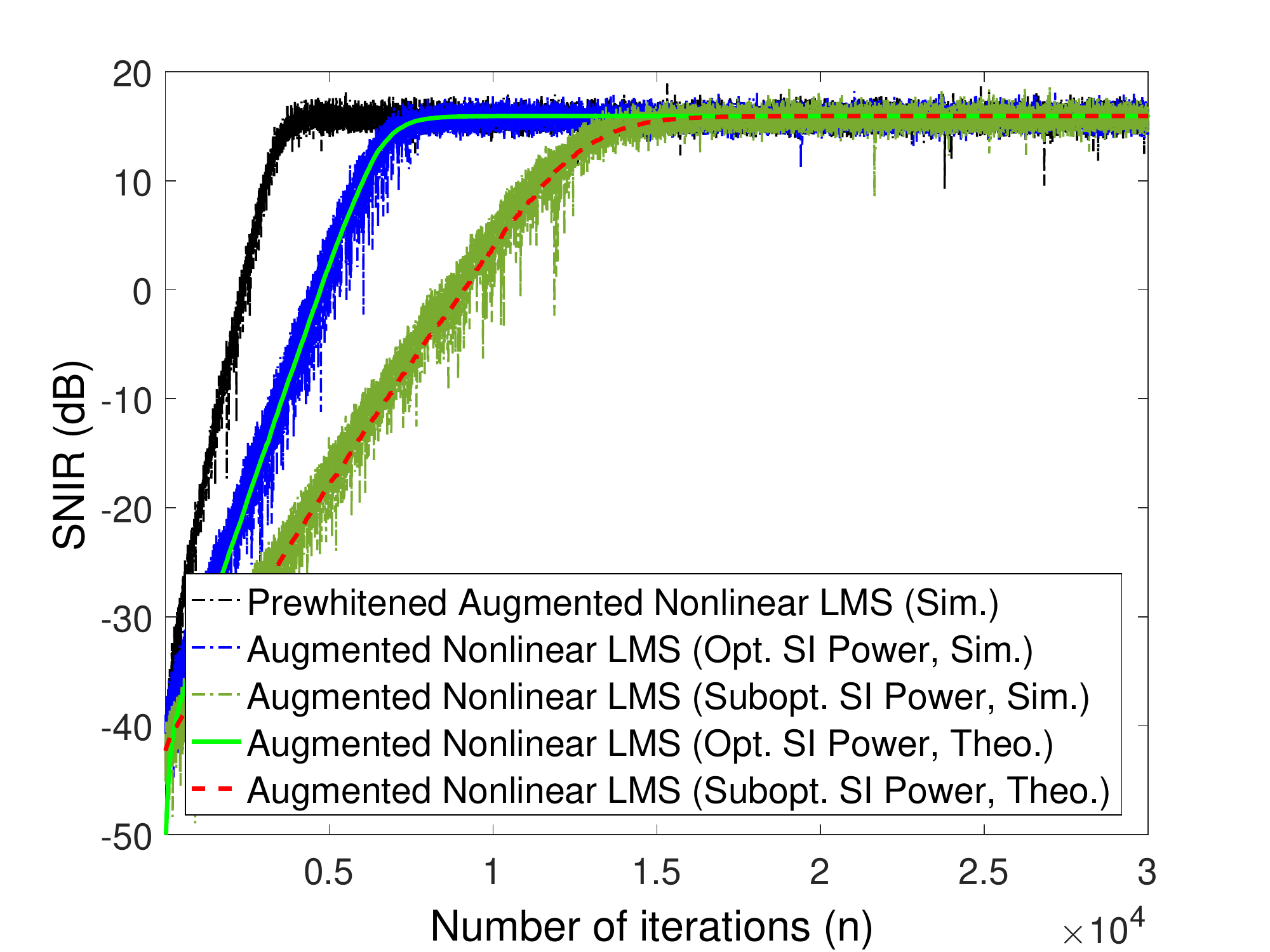} %
	\end{minipage}} \caption{Convergence speed analysis of the proposed augmented nonlinear LMS. (a) The variations of the condition number of ${{\textbf{R}}^{b}}$ against the SI power $\sigma_{x}^2$ and the Tx IQ mixer gain $k_{\rm TIQ}$. (b) Performance comparison between the theoretical and simulated SINR of augmented nonlinear LMS and the simulated SINR of its prewhitened counterpart for a Type 2 FD transceiver, with ${\mu}=0.005\mu_{{\rm {max_2}}}$.}
		\label{ANCLMSvspANCLMS}
	\end{figure*}
\vspace{0.2cm}
In order to validate performance advantages of the proposed  augmented nonlinear LMS based digital SI canceller over the conventional  augmented LMS one for full-duplex direct-conversion transceivers  in the presence of PA nonlinear distortion and frequency-dependent IQ imbalance, simulations were conducted in the MATLAB programming environment. The simulated waveforms of the transmit SI $x(n)$ and the received signal of interest $x_{{\rm {SOI}}}(n)$ were both considered to be generated from OFDM transmission systems compliant with the wireless LAN (WLAN) 802.11 standards. The numbers of subcarriers and null subcarriers of the WLAN-OFDM transmission system were respectively $K$ = 64 and $K_{{\rm {null}}}=14$. The length of cyclic prefix was $K_{{\rm {cp}}}$ = 16, the oversampling factor was $K_{{\rm {os}}}=4$, and the waveform bandwidth was $B_{{\rm {c}}}$ = 20 MHz, to give an OFDM symbol duration $T_{{\rm {sym}}}=(K+K_{{\rm {cp}}})/B_{{\rm {c}}}=4$ $\mu s$. The 16-QAM constellation scheme was used in each subcarrier. The residual
analog cancellation error was subject to a 3-tap static Rayleigh distribution,
whose detailed power-delay-profile is provided in \cite{Duarte2012}.
The frequency-dependent transmitter and receiver I/Q imbalance impulse responses
were both modeled as 2-tap static FIR filters \cite{Anttila_2008}.
In this way, the length of the end-to-end channel impulse responses for the SI vector ${\textbf{x}}(n)$
and its image component ${\textbf{x}}^{*}(n)$, that is, ${{\textbf{h}}^{\textmd{o}}}$
and ${{\textbf{g}}^{\textmd{o}}}$, was fixed to $M=5$, while that
for the IMD components ${\textbf{x}}_{{\rm {IMD}}}(n)$ and ${\textbf{x}}_{{\rm {IMD}}}^{*}(n)$,
that is, ${{\textbf{h}}_{{\rm {IMD}}}^{\textmd{o}}}$ and ${{\textbf{g}}_{{\rm {IMD}}}^{\textmd{o}}}$,
was fixed to $N=4$. All the simulation results were obtained by averaging
200 independent trials.

As stated in \textit{Remark 1}, for high Tx powers,  the conventional augmented LMS based SI
canceller yields an unavoidable steady-state bias on the estimation
of $2N$ out of $2M$ entries of the augmented end-to-end
system impulse responses $\textbf{w}^{a\textmd{o}}=[\textbf{h}^{\textmd{o}T}, \textbf{g}^{\textmd{o}T}]^{T}$,
due to the arbitrary negligence of the IMD SI components, and this
bias is quantified by \eqref{wevInfty}. This analysis is supported
by Fig. \ref{meanbehaviorfig:ACLMS}, in which the evolution of two
representative weight error coefficients, that is, ${{\widetilde{h}}_{1}}(n)$
and ${{\widetilde{h}}_{2}}(n)$, are provided. A Type 2 FD transceiver with
a transmit power at 25 dBm was considered, and two step-sizes
$\mu=0.05\mu_{{\rm {max_1}}}$ and $\mu=0.1\mu_{{\rm {max_1}}}$ were
used in  augmented LMS, where $\mu_{{\rm {max_1}}}$ is the upper bound of the step-size $\mu$, which guarantees both the mean
and mean square stability of  augmented LMS, evaluated by using \eqref{stepSizeMeanSqCvg}. Observe that the two weight
error coefficients, ${{\widetilde{h}}_{1}}(n)$ and ${{\widetilde{h}}_{2}}(n)$ converged to their theoretical steady-state biased values, rather than $0$. However, as discussed in Section \ref{ANCLMS}, due to the appropriate model fitting,
the proposed  augmented nonlinear LMS based SI canceller was able to remove this bias, as illustrated in Fig. \ref{meanbehaviorfig:ANCLMS}. By comparing the simulation results in Fig. \ref{meanbehaviorfig:ACLMS} and Fig. \ref{meanbehaviorfig:ANCLMS}, there was a cost in convergence paid by the proposed augmented nonlinear LMS in order to achieve unbiased nonlinear SI cancelleation. This is because the higher order IMD components $\textbf{x}_{\rm{IMD}}(n)$ and $\textbf{x}_{\rm{IMD}}^{*}(n)$ were considered within its underlying estimation framework, and consequently, a higher eigenvalue spread occurred in its input covariance matrix ${\textbf{R}}^b$, as compared with its linear counterpart.

We next validated the proposed mean square analysis
of both  augmented LMS and  augmented nonlinear LMS based SI cancellers. Fig. \ref{DigAtt_SINR:SINR} illustrates
the theoretical and simulated achievable SINRs in the steady-state
stage, with $\mu_{{\rm {max_1}}}$ and $\mu=0.1\mu_{{\rm {max_1}}}$
and against different levels of transmit powers for a Type 2 FD transceiver.
Observe that the empirical results were closely matched with the
analytical ones, evaluated by using \eqref{SINRAchvHighTxEqua} and
\eqref{SINRAchvANCLMSEqua} respectively for  augmented LMS and  augmented nonlinear LMS. These also conform with the analysis in \textit{Remark 3} and \textit{Remark
5} which states that a smaller step-size $\mu$ enables a better steady-state SINR
performance for both SI cancellers, but at the cost of slower convergence.
Fig. \ref{DigAtt_SINR:SINR} also justifies the motivations to propose
the  augmented nonlinear LMS based SI canceller in the sense that its SINR performance
is much better than that of  augmented LMS in the high transmit power range
where the IMD SI component ${{\textbf{x}}_{{\rm {IMD}}}(n)}$ and
its image counterpart become dominant, since both components have been
generically considered as a part of the augmented nonlinear input vector within the proposed
 augmented nonlinear LMS. As expected, when the transmit power was low enough, both
SI cancellers provided a nearly identical SINR performance, since
the higher order components became negligible. This also resulted in
the theoretical SINR performance of  augmented LMS in \eqref{SINRAchvHighTxEqua},
derived for a high transmit power, asymptotically converge to that
in \eqref{SINRAchv}. The above discussion is also applicable for Fig.
\ref{DigAtt_SINR:DigAtt}, where we compared the digital attenuation
capability of the two considered SI cancellers after convergence with
a step-size $\mu=0.05\mu_{{\rm {max_1}}}$, for both Type 1 and Type
2 FD transceivers and against different levels of transmit powers. The digital
attenuation performance is a measure of the amount of SI before and
after applying an SI canceller, defined as the power ratio between
the desired signal $d(n)$ and the corresponding estimation error
of an SI canceller \cite{Korpi2014}. Observe the excellent agreement
between the simulated results and their theoretical evaluations, as
well as the performance advantages of the proposed  augmented nonlinear LMS over the
conventional  augmented LMS in the high transmit power range.
%
\vspace{0.2cm}
\subsection{Further Convergence Speed Improvement of the Proposed  Augmented Nonlinear LMS based SI Canceller}\label{sec:prewhtn}
\vspace{0.2cm}%
When higher order IMD components $\textbf{x}_{{\rm {IMD}}}(n)$ and $\textbf{x}_{{\rm {IMD}}}^{*}(n)$ are strong in a FD transceiver, a potential drawback encountered by adaptive SI cancellers lies in the slow convergence, incurred by the high eigenvalue spread within the input covariance matrix ${{\textbf{R}}^{b}}$ \cite{Anttila2013}. In fact, based on \eqref{eigenvalue}, the condition number of ${{\textbf{R}}^{b}}$, the ratio between its maximum and minimum eigenvalues, is denoted by ${\mathcal C}_{{{\textbf{R}}^{b}}}$, and can be calculated as
\begin{align}\label{CondNo}
{\mathcal C}_{{{\textbf{R}}^{b}}}\!=\!\frac{{\lambda_{2}^{b}}}{{\lambda_{3}^{b}}}\!=\!\frac{1\!+\!6{k_{{\rm {TIQ}}}^{3}}\sigma_{x}^{4}\!+\!\sqrt{{1\!-\!2{k_{{\rm {TIQ}}}^{3}}\sigma_{x}^{4}}\!+\!36{k_{{\rm {TIQ}}}^{6}}\sigma_{x}^{8}}}{1\!+\!6{k_{{\rm {TIQ}}}^{3}}\sigma_{x}^{4}\!-\!\sqrt{{1\!-\!2{k_{{\rm {TIQ}}}^{3}}\sigma_{x}^{4}}\!+\!36{k_{{\rm {TIQ}}}^{6}}\sigma_{x}^{8}}}
\end{align}
This reveals that the value of ${\mathcal C}_{{{\textbf{R}}^{b}}}$ depends on both the Tx mixer gain, ${k_{{\rm {TIQ}}}}$, and the SI power, $\sigma_{x}^2$. Therefore, for the FD transceiver with a given ${k_{{\rm {TIQ}}}}$, one possible solution to increase the convergence speed of the augmented nonlinear LMS is to synthetically scale $\sigma_{x}^2$, so that, the minimum of ${\mathcal C}_{{{\textbf{R}}^{b}}}$ can be achieved. In fact, as first proven in Appendix B and further illustrated in Fig. \ref{ANCLMSvspANCLMS:CondNo}, although the global minimum of ${\mathcal C}_{{{\textbf{R}}^{b}}}$ does exist, it is greater than unity. Owing to the closed-loop structure of the FD transceiver, the knowledge of the augmented nonlinear SI vector ${{\textbf{x}}^{b}}(n)$ is ideally inherent in the receiver end. Therefore, a more efficient solution can be established by using standard pre-whitening to decompose ${{\textbf{R}}^{b}}$ as ${{\textbf{R}}^{b}}={\bf {U}}{{\bm {\Lambda}}^{b}}{\bf {U}}^{H}$ and consequently to produce a whitened input vector, ${\widetilde{{\textbf{x}}}^{b}}(n)={\bm{\Phi}}{{\textbf{x}}^{b}}(n)$, where ${\bm{\Phi}}=({{\bm {\Lambda}}^{b}})^{-\frac{1}{2}}{\bf {U}}^{H}$. In this way, the speed of gradient decent of augmented nonlinear LMS is fixed and normalized, i.e., with a unity condition number, thus facilitating practical applications. This is supported by Fig. \ref{ANCLMSvspANCLMS:cvg-state}, which shows both the theoretical and simulated convergence behavior, measured in terms of SINR, of the proposed augmented nonlinear LMS based SI canceller and its data-whitening assisted version for a Type 2 FD transceiver with a Tx power at 15 dBm and a Tx IQ mixer gain $k_{\rm TIQ}=6$ dB.  The step-size was ${\mu}=0.005\mu_{{\rm max_2}}$, where $\mu_{{\rm {max_2}}}$ is its upper bound on $\mu$ which guarantees both the mean and mean square stability of augmented nonlinear LMS, evaluated by using \eqref{stepSizeMeanCvgJ} and \eqref{Mean_Square_ANCLMS}, respectively. In this case, the optimal SI power $\sigma_{x}^2$ was roughly -13 dBm for the minimum eigenvalue spread. The simulation results for a suboptimal SI power, $\sigma_{x}^2=-10$ dBm, are also provided. We observe that the proposed theoretical evaluation accurately described the empirical SINR evolution of augmented nonlinear LMS in both the transient and steady-state stages. We also observe that, with the optimal SI power, the augmented nonlinear LMS exhibited a faster convergence, about 7000 iterations, as compared with the considered suboptimal case. However, when the prewhitening scheme was employed, it merely required about 3000 iterations to arrive at the steady state.

\section{Conclusion}
\label{con} The impact of typical front-end non-idealities on a future 5G systems, such as nonlinear PA distortion, IQ imbalance, thermal noise and quantization noise, has been rigorously analyzed over both the mean and mean square performances of the conventional augmented LMS based digital SI canceller for fully-duplex direct-conversion transceivers. We have here quantified the estimation bias and the second order performance suboptimality exhibited by the augmented LMS in case of high transmit powers. To rectify these drawbacks, an augmented nonlinear LMS based SI canceller has been proposed, which naturally accounts for those higher order components by virtue of a widely nonlinear model fit. Its performance advantages over augmented LMS have been verified both theoretically and through numerical validation. To further increase convergence speed of the proposed scheme, the standard data pre-whitening scheme has also been employed in this context. Illustrative simulations on two representative types of FD transceivers for OFDM-based WLAN standard compliant waveforms support the analysis.
\vspace{-0.3cm}
\begin{appendices}%
\section{Evaluation of the term ${\rm {diag}}\{{{{\textbf{Q}}_{3}}(\infty)}\}$
in the case of high transmit power}\label{App:AppendixA}
\setcounter{equation}{57}
Based on \eqref{un_vec} and \eqref{Qs}, the diagonal elements of ${{{\textbf{Q}}_{3}}(n)}$ can be derived as
\begin{align*}
&{\rm {diag}}\{E[{{{\textbf{Q}}_{3}}(n)}]\}\nonumber\\
&={\rm {diag}}\{E[{\textbf{x}}_{{\rm {IMD}}}^{T}(n){{\textbf{h}}_{{\rm {IMD}}}^{\textmd{o}}}(n){\textbf{x}}^{a*}(n)]E[{{\widetilde{\textbf{w}}}^{aH}}(n)]\}\nonumber\\
&+{\rm {diag}}\{E[{\textbf{x}}_{{\rm {IMD}}}^{H}(n){{\textbf{g}}_{{\rm {IMD}}}^{\textmd{o}}}(n){\textbf{x}}^{a*}(n)]E[{{\widetilde{\textbf{w}}}^{aH}}(n)]\}\nonumber\\
&+{\rm {diag}}\{E[v(n){\textbf{x}}^{a*}(n)]E[{{\widetilde{\textbf{w}}}^{aH}}(n)]\} \label{diagE3Compute}
\end{align*}
while the standard independence assumptions yield
\begin{equation}
{\rm {diag}}\{E[v(n){\textbf{x}}^{a*}(n)]E[{{\widetilde{\textbf{w}}}^{aH}}(n)]\}={\textbf{0}}\label{diagvnxanwhn}
\end{equation}
Now, based on \eqref{Eunxaconjn} and \eqref{diagvnxanwhn}, we have
\begin{align*}
& {\rm {diag}}\{{{{E[\textbf{Q}}_{3}}(n)]}\}\!=\![{h_{{\rm {IMD}},1}}k_{{\rm {TIQ}}}^{3/2}E[{\left|{x(n)}\right|^{4}}]E[\widetilde{w}_{1}^{*}(n)],\nonumber\\
& {h_{{\rm {IMD}},2}}k_{{\rm {TIQ}}}^{3/2}E[{\left|{x(n-1)}\right|^{4}}]E[\widetilde{w}_{2}^{*}(n)], \ldots,\nonumber\\
& {h_{{\rm {IMD}},N}}k_{{\rm {TIQ}}}^{3/2}E[{\left|{x(n-N+1)}\right|^{4}}]E[\widetilde{w}_{N}^{*}(n)], {\textbf{0}}_{M-N}^{T},\nonumber\\
& {g_{{\rm {IMD}},1}}k_{{\rm {TIQ}}}^{3/2}E[{\left|{x(n)}\right|^{4}}]E[\widetilde{w}_{1}^{*}(n)], \ldots,\nonumber\\
& {g_{{\rm {IMD}},N}}k_{{\rm {TIQ}}}^{3/2}E[{\left|{x(n)}\right|^{4}}]E[\widetilde{w}_{N}^{*}(n)], {\textbf{0}}_{M-N}^{T}]
\label{detailed Q3}
\end{align*}
Note that $E[{\left|{x(n)}\right|^{4}}]=2\sigma_{x}^{4}$, and hence, the steady-state evaluation of ${\rm {diag}}\{E[{{{\textbf{Q}}_{3}}(n)}]\}$,
that is, ${\rm {diag}}\{E[{{{\textbf{Q}}_{3}}(\infty)}]\}$, is now
subject to $E[{\widetilde{\textbf{w}}}^{a}(\infty)]$, and based on
\eqref{wevInfty}, this gives
\begin{equation*}
{\rm {diag}}\{{{{\textbf{Q}}_{3}}(\infty)}\}=4k_{{\rm {TIQ}}}^{3}\sigma_{x}^{6}{{\textbf{p}}_{{\rm {IMD}}}^{\textmd{o}}}\label{diagE3}
\end{equation*}
where ${{\textbf{p}}_{{\rm {IMD}}}^{\textmd{o}}}$ is defined as
\begin{align}
{{\textbf{p}}_{{\rm {IMD}}}^{\textmd{o}}} &=[{\left|{h_{{\rm {IMD}},1}}\right|^{2}},{\left|{h_{{\rm {IMD}},2}}\right|^{2}},\ldots,{\left|{h_{{\rm {IMD}},N}}\right|^{2}},{\textbf{0}}_{M-N}^{T},\nonumber\\
& {\left|{g_{{\rm {IMD}},1}}\right|^{2}},{\left|{g_{{\rm {IMD}},2}}\right|^{2}},\ldots,{\left|{g_{{\rm {IMD}},N}}\right|^{2}},{\textbf{0}}_{M-N}^{T}]^{T}
\label{phIMD}
\end{align}
Also note that $\Re[{{{\textbf{Q}}_{3}}(\infty)]}={{{\textbf{Q}}_{3}}(\infty)}$,
since it is real-valued. 

\section{Minimization of ${\mathcal C}_{{{\textbf{R}}^{b}}}$}\label{App:AppendixB}
By defining a positive variable $\epsilon={k_{{\rm {TIQ}}}^{3}}\sigma_{x}^{4}$, the condition number ${\mathcal C}_{{{\textbf{R}}^{b}}}$ in \eqref{CondNo} becomes
\begin{equation*}\label{CRb}
{\mathcal C}_{{{\textbf{R}}^{b}}}=
\frac{1 + 6\epsilon + \sqrt {1 - 2\epsilon + 36{\epsilon^2}}}{1 + 6\epsilon - \sqrt {1 - 2\epsilon + 36{\epsilon^2}} }
\end{equation*}
After some mathematical manipulations, its first derivative with respect to $\epsilon$ can be expressed as
\begin{align*}\label{dCRb}
\frac{\partial{\mathcal C}_{{\textbf{R}}^{b}}}{\partial\epsilon} &=\frac{14(6\epsilon - 1)}{\left(1 + 6\epsilon - \sqrt {1 - 2\epsilon + 36{\epsilon^2}}\right)^2\sqrt {1 - 2\epsilon + 36{\epsilon^2}}}
\end{align*}
It can be easily verified that this function has a zero only at $\epsilon=\frac{1}{6}$, and its denominator is always positive for $\epsilon >0$. Therefore, for $\epsilon \in (0,\frac{1}{6})$, we have ${\partial{\mathcal C}_{{\textbf{R}}^{b}}}/{\partial\epsilon}<0$, while for $\epsilon \in (\frac{1}{6},\infty)$, we have ${\partial{\mathcal C}_{{\textbf{R}}^{b}}}/{\partial\epsilon}>0$, indicating that the global minimum of ${\mathcal C}_{{{\textbf{R}}^{b}}}$ exists at $\epsilon=\frac{1}{6}$. After some algebraic manipulations, the minimum condition number is found to be $\frac{17+4\sqrt{15}}{7} \approx 4.64$.
\end{appendices}

{\small{} \bibliographystyle{IEEEtran}
\bibliography{Manuscript}
 }{\small \par}

\end{document}